%% file: paper.tex
\documentclass[conference]{IEEEtran}
\IEEEoverridecommandlockouts
\usepackage{cite}
\usepackage{amsmath,amssymb,amsfonts}
\usepackage{algorithmic, setspace}
\usepackage{textcomp}
\usepackage{xcolor}
\usepackage{color}
\usepackage{url}
\newtheorem{definition}{Definition}
\usepackage[T1]{fontenc}
\usepackage{array}
\usepackage{calc}
\usepackage{multirow}
\usepackage{graphicx, float}
\usepackage{subfigure}
\usepackage[linesnumbered,ruled]{algorithm2e}

\usepackage{makecell}
\def\BibTeX{{\rm B\kern-.05em{\sc i\kern-.025em b}\kern-.08em
    T\kern-.1667em\lower.7ex\hbox{E}\kern-.125emX}}
\input{annotations}

\begin{document}

\title{Self-Checking Deep Neural Networks in Deployment
}


\author{\IEEEauthorblockN{Yan Xiao\IEEEauthorrefmark{1}, Ivan Beschastnikh\IEEEauthorrefmark{2}, David S. Rosenblum\IEEEauthorrefmark{3}, 
		Changsheng Sun\IEEEauthorrefmark{1}, Sebastian Elbaum\IEEEauthorrefmark{5}, \\Yun Lin\IEEEauthorrefmark{1} and Jin Song Dong\IEEEauthorrefmark{1}} \IEEEauthorblockA{\IEEEauthorrefmark{1}School of Computing, National University of Singapore, Singapore\\
    dcsxan@nus.edu.sg, changsheng\_sun@outlook.com, dcsliny@nus.edu.sg, dcsdjs@nus.edu.sg} 
	\IEEEauthorblockA{\IEEEauthorrefmark{2}Department of Computer Science, University of British Columbia, Vancouver, BC, Canada, 
	  bestchai@cs.ubc.ca}
	\IEEEauthorblockA{\IEEEauthorrefmark{3}Department of Computer Science, George Mason University, Fairfax, VA, USA,
	  dsr@gmu.edu} 
	\IEEEauthorblockA{\IEEEauthorrefmark{5}Department of Computer Science, University of Virginia, Charlottesville, VA, USA,
	  selbaum@virginia.edu}}

\maketitle
\thispagestyle{plain}
\pagestyle{plain}

\begin{abstract}
The widespread adoption of Deep Neural Networks (DNNs) in important
domains raises questions about the trustworthiness of DNN
outputs. Even a highly accurate DNN will make mistakes some of the
time, and in settings like self-driving vehicles these mistakes must
be quickly detected and properly dealt with \emph{in deployment}.

Just as our community has developed effective techniques and
mechanisms to monitor and check programmed components, we believe it
is now necessary to do the same for DNNs.
In this paper we present \textit{DNN self-checking} as a
process by which internal DNN layer features are used to check DNN
predictions. We detail \emph{SelfChecker}, a self-checking system that
monitors DNN outputs and triggers an alarm if the internal layer
features of the model are inconsistent with the final
prediction. SelfChecker also provides \emph{advice} in the form of an
alternative prediction.

We evaluated SelfChecker on four popular image datasets and three DNN
models and found that SelfChecker triggers correct alarms on 60.56\%
of wrong DNN predictions, and false alarms on 2.04\% of correct DNN
predictions. This is a substantial improvement over prior work
(\textsc{SelfOracle}, \textsc{Dissector}, and ConfidNet).
In experiments with self-driving car scenarios, SelfChecker triggers
more correct alarms than \textsc{SelfOracle} for two DNN models
(DAVE-2 and Chauffeur) with comparable false alarms. Our
implementation is available as open source.
\end{abstract}


\input{body}

\bibliographystyle{IEEEtran}
\bibliography{paper}

\end{document}

%% file: annotations.tex
\usepackage{ifthen}
\usepackage[normalem]{ulem} 
\usepackage{xcolor}
\usepackage{amssymb}

\newboolean{showedits}
\setboolean{showedits}{true} 
\ifthenelse{\boolean{showedits}}
{
	\newcommand{\del}[1]{\textcolor{red}{\sout{#1}}} 
}{
	\newcommand{\del}[1]{} 
	
}

\newboolean{showcomments}
\setboolean{showcomments}{true} 
\newcommand{\id}[1]{$-$Id: scgPaper.tex 32478 2010-04-29 09:11:32Z oscar $-$}

\ifthenelse{\boolean{showcomments}}
{\newcommand{\nbc}[3]{
 {\colorbox{#3}{\bfseries\sffamily\scriptsize\textcolor{white}{#1}}}
 {\textcolor{#3}{\sf\small$\blacktriangleright$\textit{#2}$\blacktriangleleft$}}}
 }
{\newcommand{\nbc}[3]{}
 \renewcommand{\del}[1]{} 
 }

\definecolor{ibcolor}{rgb}{0.9,0.5,0}
\definecolor{dsrcolor}{rgb}{0.5,0.6,0}
\definecolor{cfcolor}{rgb}{0,0.5,0.9}
\definecolor{oldcolor}{rgb}{0.2,0.2,0.2}
\definecolor{tdcolor}{rgb}{1.0,0,0}

\newcommand{\xy}[1]{{\color{black} #1}}

%% file: body.tex
\section{Introduction}

Deep Neural Networks (DNNs) are now used in a variety of domains, including speech processing~\cite{hinton2012deep}, NLP~\cite{sutskever2014sequence}, medical diagnostics~\cite{ciresan2012deep}, image processing~\cite{ciregan2012multi}, robotics~\cite{zhang2015towards} and even reconstruction of brain circuits~\cite{helmstaedter2013connectomic}. The power and accuracy of DNNs have led to deployments of Deep Learning (DL) systems in safety- and security-critical domains, including self-driving cars~\cite{bojarski2016end}, malware detection~\cite{yuan2014droid} and aircraft collision avoidance systems~\cite{julian2016policy}.  
Such domains have a low tolerance for mistakes. The software systems in a self-driving car, for example, must have high assurance in deployment.

Unfortunately, the stochastic nature of DL virtually ensures that DL models will not achieve 100\% accuracy, even on the training dataset. Since in mission-critical applications a wrong DNN decision could be costly, we believe that such applications must include logic to (1)~\emph{check} the trustworthiness of a DNN's output, and (2)~raise an \emph{alarm} when there is low confidence in the output.
Our community has developed such methods for programmed components~\cite{chimento2015s,mitsch2016modelplex,lin2017feedback} and now is the time to do so for learned ones like DNNs.

Trustworthiness of a simple DNN can be measured with softmax probabilities~\cite{hendrycks2016baseline}, or information theoretic metrics, such as entropy~\cite{steinhardt2016unsupervised} and mutual information~\cite{shannon1948mathematical}.
However, in complex DNNs with many layers and neurons, softmax probabilities and entropy are unreliable confidence estimators of the prediction~\cite{vasudevan2019towards,corbiere2019addressing}. Even for abnormal samples, DNNs may still produce overconfident posterior probabilities.
For example, when we built classifiers for VGG-16~\cite{simonyan2014very} on CIFAR-10~\cite{krizhevsky2009learning}, we found that 75\% of predictions that were incorrect had maximum softmax probabilities over 70\%; and 63\% incorrect predictions had maximum softmax probabilities over 80\%.  
We had similar results on other datasets and models. This illustrates the unreliability of the softmax probabilities as confidence estimators of the final prediction.

Our goal is to build a general-purpose system that monitors a deployed DNN's predictions during inference, raises an \emph{alarm} if there is low confidence in the predictions, and provides an alternative prediction that we call an \emph{advice}.
A key challenge in building such a system is finding a source of additional information to check DNN outputs. The inspiration for our work comes from Kaya et al., who study internal DNN behavior~\cite{kaya2019shallow}. They found that a DNN can reach a correct prediction \emph{before} the final layer. In fact, the final layer of a DNN may change a correct internal prediction into an incorrect prediction. This work illustrates that features extracted from internal layers of a DNN contain information that can be used to cross-check a model's output.

Inspired by Kaya et al.'s work, we define \textit{self-checking} as a process by which internal DNN layer features are used to check DNN predictions. In this paper we describe a novel self-checking system, called \textbf{SelfChecker}, that triggers an alarm if the internal layer features of the model are inconsistent with the final prediction. SelfChecker also provides \emph{advice} in the form of an alternative prediction. SelfChecker assumes that the training and validation datasets come from a distribution similar to that of the inputs that the DNN model will face in deployment.

SelfChecker uses kernel density estimation (KDE) to extrapolate the probability density distributions of each layer's output by evaluating the DNN on the training data.
Based on these distributions, the density probability of each layer's outputs can be inferred when the DNN is given a test instance.
SelfChecker measures how the layer features of the test instance are similar to the samples in the training set. If a majority of the layers indicate inferred classes that are different from the model prediction, then SelfChecker triggers an alarm.
In addition, not all layers can contribute positively to the final prediction~\cite{kaya2019shallow}. SelfChecker therefore uses a search-based optimization to select a set of optimal layers to generate a high quality alarm and advice.

We evaluated SelfChecker's alarm and advice mechanisms with experiments on four popular and publicly-available data\-sets (MNIST, FMNIST, CIFAR-10, and CIFAR-100) and three DNNs (ConvNet, VGG-16, and ResNet-20) against three competing approaches (\textsc{SelfOracle}~\cite{stocco2020misbehaviour}, \textsc{Dissector}~\cite{wang2020dissector}, and ConfidNet~\cite{corbiere2019addressing}). Our results show that SelfChecker achieves the highest F1-score (68.07\%), which is 8.77\% higher than the next best approach (ConfidNet). Our evaluation of SelfChecker's DNN prediction checking runtime shows an acceptable time overhead of 34.98ms.
We also compared SelfChecker to the state-of-the-art approach for self-driving car scenarios (\textsc{SelfOracle}~\cite{stocco2020misbehaviour}), and found that SelfChecker triggers more correct alarms and a comparable number of false alarms.

Our paper makes the following three contributions:
\begin{itemize}
\renewcommand{\labelitemi}{$\star$}
	\item We present the design of SelfChecker, which uses density distributions of layer features and a search-based layer selection strategy to trigger an alarm if a DNN model output has low confidence. We show that SelfChecker achieves better alarm accuracy than previous work.
	\item Unlike existing work, SelfChecker provides \emph{advice} in the form of an alternative prediction. We find that models on a 10-class dataset can use this advice to achieve higher prediction accuracy. 
	\item We demonstrate the effectiveness of SelfChecker's alarms and advice on publicly available DNNs, ranging from small models (ConvNet) to large and complex models (VGG-16 and ResNet-20), and self-driving car scenarios. Our implementation is open-source\footnote{https://github.com/self-checker/SelfChecker}.
\end{itemize}

\section{Background and Motivation} \label{sec:motivation}

In a deep neural network (DNN), an input is fed into the input layer, then passed through a series of hidden layers that extract features from the input using activation functions attached to neurons, and the process concludes with the output layer, which uses the extracted features to output a prediction using either \emph{classification} (from a categorical set of classes) or \emph{regression} (in the form of real-valued ordinals). 
The behavior of a layer during inference thus can be characterized by its vector of neuron activation outputs.   In what follows, we refer to these layer-wise vectors of activation outputs as the \emph{layer features} analyzed by our approach.

\subsection{The Promise of Using Layer Features}

DNNs make decisions based on features extracted from training data. But how can we judge if a model is making a wrong decision for a given test instance? One way is to check whether the model has previously observed a similar instance during training. 
This raises the question of how to define the similarity between a test instance $\mathbf{x}$ and a training instance $\mathbf{x}^\prime$. Most existing studies use a distance-based measure~\cite{sun2018testing}, such as $L_{p}$ or cosine similarity.
We think this is problematic since the inputs are complex enough and need DNNs to extract features, so we doubt that a distance measure defined directly on the inputs can properly capture similarity. 

\textbf{Instead, we use the features of the inputs extracted by internal layers in DNNs to capture similarity.} Specifically, we define the similarity as the likelihood of the DNN having seen a similar layer features during training. We use probability density distributions extrapolated from the training process to measure the similarity between layer features of a given input and those observed for training data.

\begin{figure}[h]
	\centering	
	\begin{minipage}[c]{1\linewidth}		
		\centering		
		\includegraphics[width=1\linewidth]{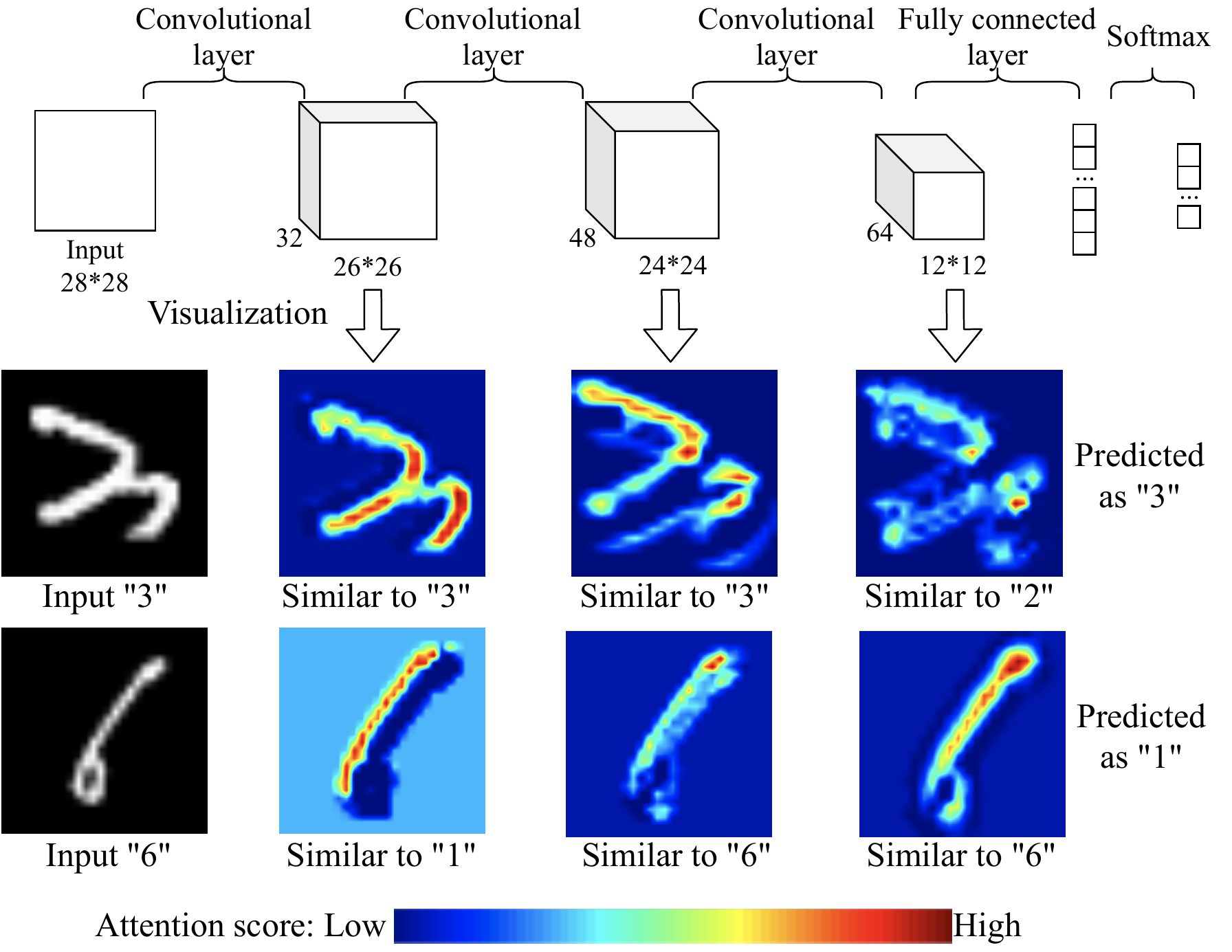}		
	\end{minipage}%
	
	\caption{
		The top of the figure depicts the architecture of a Convolutional Neural Network with three convolutional layers to classify digit images. The two rows of images at the bottom depict attention heatmaps for the associated layers when given test inputs for digits 3 and 6, respectively.
	}
	\label{fig:motivate}
\end{figure}
Fig.~\ref{fig:motivate} presents a motivating example where a Convolutional Neural Network (CNN) with three convolutional layers trained on MNIST is used to classify images of digits 3 and 6, while outputting labels ``3'' and ``1'' as the respective predictions. To visualize \emph{where} the features of each layer focus, we apply Grad-CAM~\cite{selvaraju2017grad} to highlight the attention heatmap on the original images as shown in the bottom two rows of images in Fig.~\ref{fig:motivate}. The heatmap images show that different layers have different points of focus. For example, the first and second images of digit 3 are similar to 3 itself, but the third image is closer to digit 2. Similarly, the first image of digit 6 is similar to digit 1, but the second and third images are similar to 6.

\begin{figure*}[hbtp]
	\begin{centering}
		\includegraphics[width=0.85\linewidth]{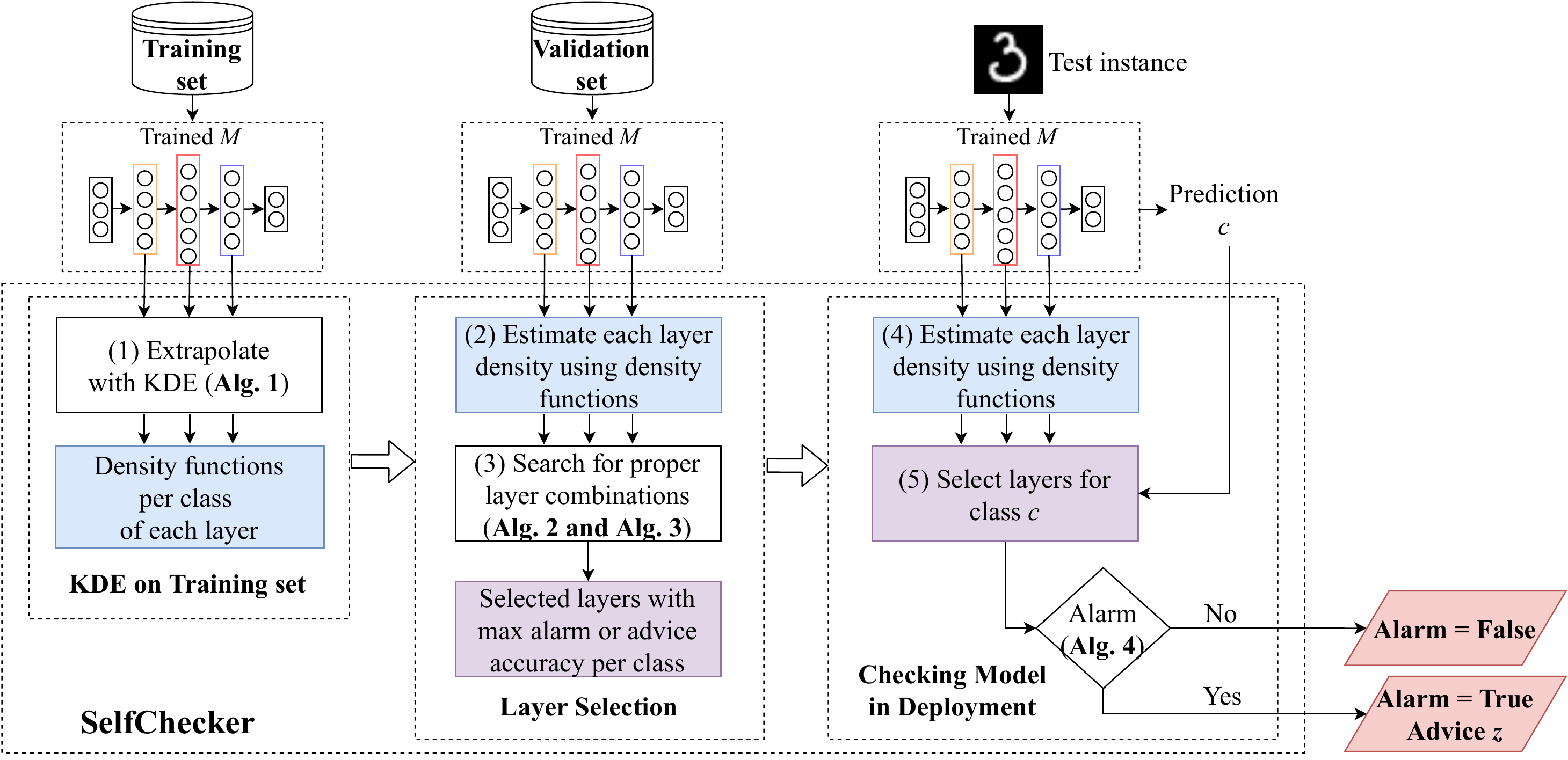}
		\par\end{centering}
	        \caption{The design of SelfChecker and its integration
                  with a trained model and model predictions.
                }
	\label{fig:framework}
\end{figure*}

Although the CNN misclassifies the second image, in both cases the images appear to be recognized correctly by one or more hidden layers. This example thus illustrates the promise of using layer features to
check the model's classification of a test instance.

DNNs exist in many variants and can be combined to form more complex
models. For example, models used in urban flow
prediction~\cite{pan2019urban, liang2019urbanfm} combine convolutional, graph and recurrent neural
nets. However, all these DNNs extract features using internal
layers, and that is the focus of our research.


The design we present targets DNN classifiers with convolutional
layers and fully-connected layers. Our system also works for
regression networks by transforming the network into a binary
classification problem. Since our design uses layer features, it
should work on other types of DNNs, such as recurrent neural networks.
We leave the evaluation of our system on other DNN types to future work.

\subsection{The Challenges of Using Layer Features}

The preceding example also raises two challenges
that a technique using layer features must resolve:

\begin{itemize}

\item Which layers should be selected for checking the classification of a test instance? For example, does selecting
  more layers lead to a better checker?
\item How should the features from the different layers be aggregated
  --- either to determine if an alarm should be raised, or to
  produce alternative advice?

\end{itemize}
Resolving these questions is the goal of this paper.

\textbf{Problem statement.} Given a trained DNN classifier and a test instance, we aim to develop a systematic method called SelfChecker for determining whether the DNN will misclassify the test instance, based on extensive checking the DNN's internal features. First, SelfChecker should trigger an \emph{alarm} if it detects a potential misclassification of the test instance. Second, and going beyond the previous studies~\cite{stocco2020misbehaviour,wang2020dissector,corbiere2019addressing}, SelfChecker should provide \emph{advice} once an alarm is triggered, in the form of an alternative classification.  Our goal is for SelfChecker to achieve high accuracy in both triggering alarms and offering advice.

\section{Design of SelfChecker}\label{sec:design}

The goals of SelfChecker are (1)~to check a DNN's prediction, (2)~to raise an \emph{alarm} if the DNN's prediction is determined to be incorrect, and (3)~to provide an \emph{advice}, or an alternative prediction.

SelfChecker's \emph{training module} is used after the model has been trained to
configure SelfChecker's behavior in deployment. The training module
uses the training and validation datasets, as well as the trained
model to generate a deployment configuration.

SelfChecker's \emph{deployment module} runs along with the inference process: it
analyses the internal features of a DNN when the model is given a test
instance and provides an alarm as well as an advice if it detects an
inconsistency in the model's output. To detect these inconsistencies,
the deployment module uses the configuration supplied to it by the
training module.

Note that although SelfChecker analyses the features extracted from
the internal layers of a DNN, the training module is independent from
the architecture of the model and requires no model modifications or
retraining. The deployment module, however, is specific to a DNN.

\begin{figure*}[ht]
	\centering
	\subfigure[Histogram]{
		\begin{minipage}[t]{0.31\linewidth}
			\centering
			\includegraphics[width=4.6cm]{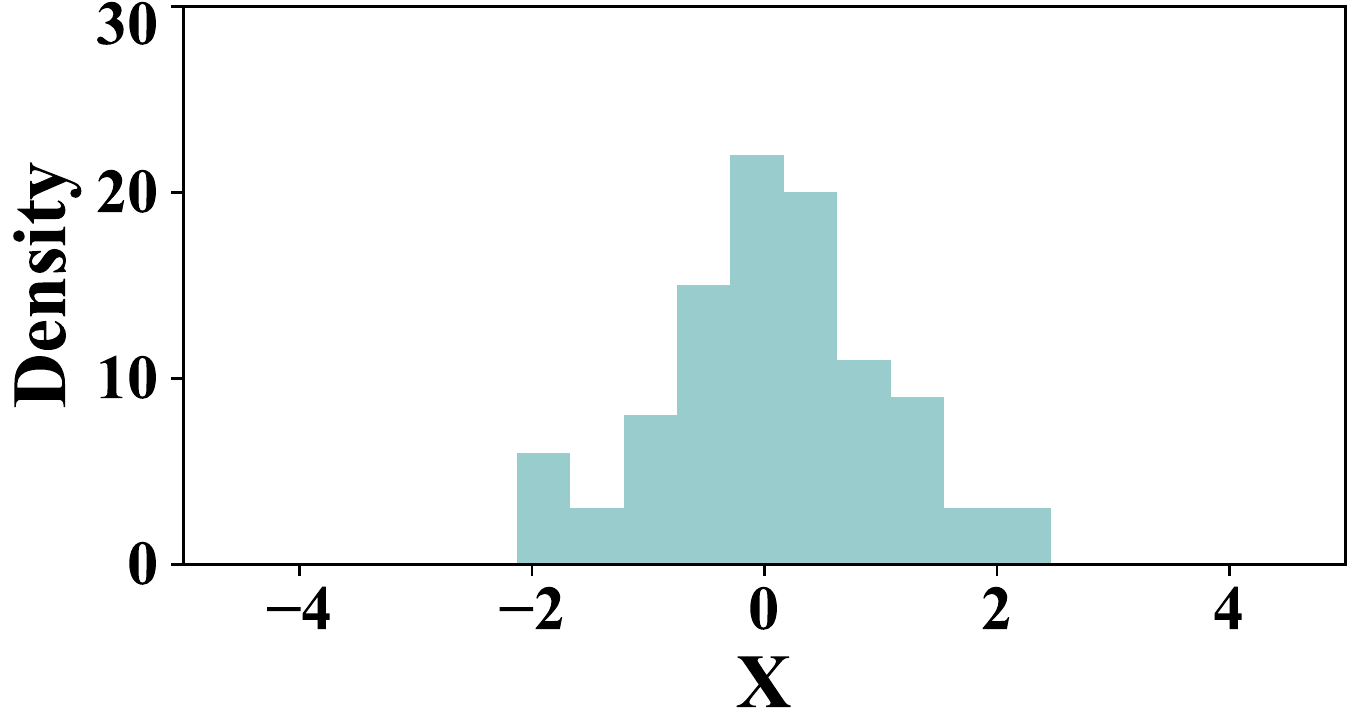} 
		\end{minipage}
	}
	\subfigure[KDE]{
		\begin{minipage}[t]{0.31\linewidth}
			\centering
			\includegraphics[width=4.6cm]{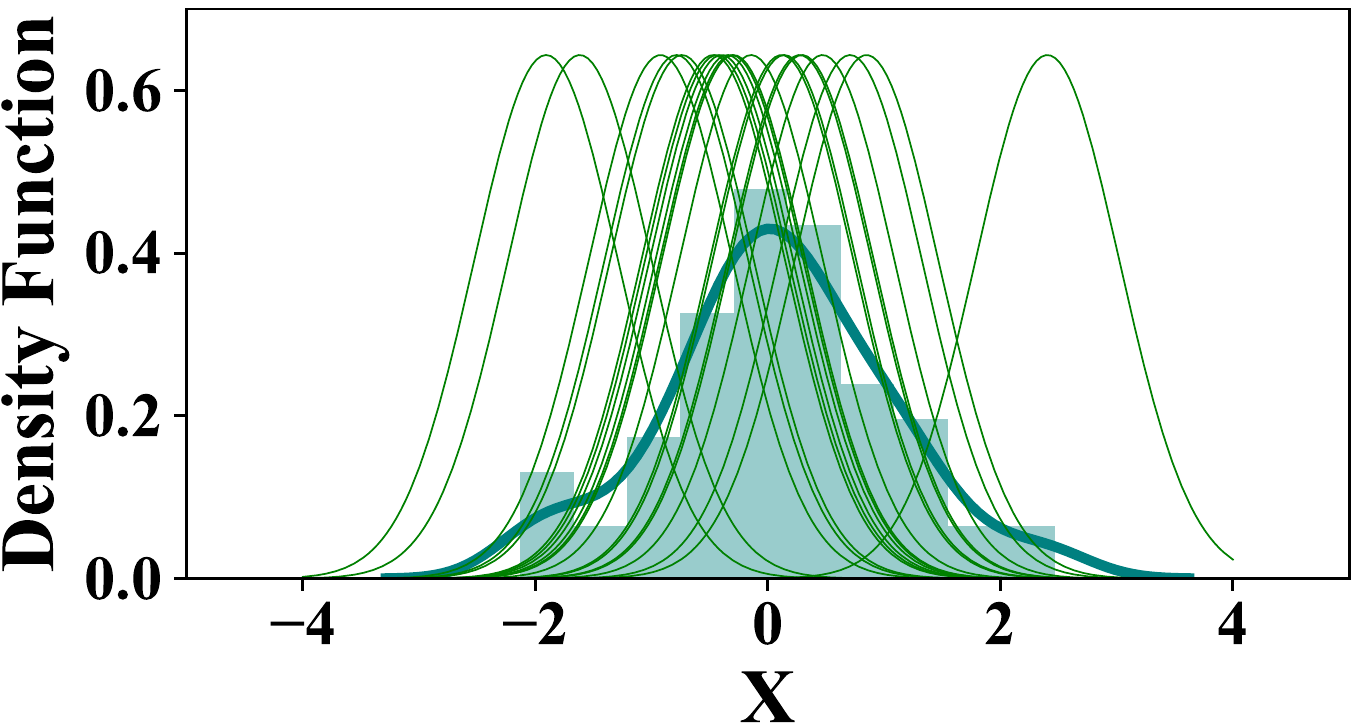} 
		\end{minipage}
	}
	\subfigure[Bandwidth]{
		\begin{minipage}[t]{0.31\linewidth}
			\centering
			\includegraphics[width=4.6cm]{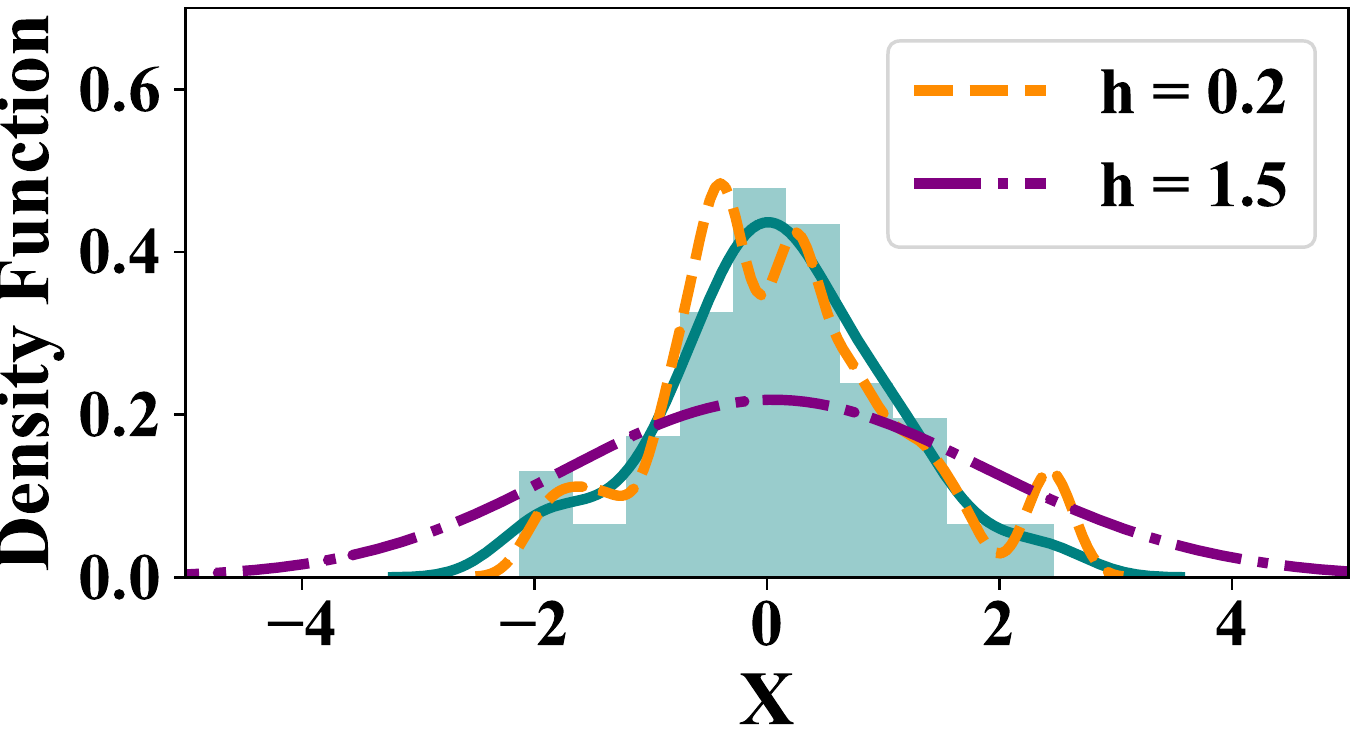} 
		\end{minipage}
	}
	\caption{An example to illustrate KDE computation with \emph{(a)} showing the set of input 1D points, \emph{(b)} showing how to obtain the distribution using a KDE, and \emph{(c)} showing the distributions obtained by using different bandwidths.
	}
	\label{fig:kde}
\end{figure*}

Fig.~\ref{fig:framework} overviews our approach. Given a DNN model \textit{M} trained on training dataset $\mathit{D_{train}}$ and validated on validation set $\mathit{D_{valid}}$, for each layer in \textit{M}, SelfChecker's training module first (1) computes layer-wise density distributions of each class using kernel density estimation (KDE)~\cite{terrell1992variable} on $\mathit{D_{train}}$ (Section~\ref{sec:kde}). Based on the distributions, (2) SelfChecker can estimate the density values of each validation or test instance on each class. The higher the values of the class, the more similar the features of the instance in this layer are to the specific class. After SelfChecker obtains all of estimated density values on $\mathit{D_{valid}}$ across all layers, SelfChecker (3) finds the optimal layer combinations to reach the best alarm and advice accuracy. Since different classes produce distinctive feature behaviors in different layers, SelfChecker uses global search to find the optimal layer combinations per class (Section~\ref{sec:search}).

Finally, when the model is presented with a test instance in deployment, SelfChecker's deployment module decides whether to provide an alarm as well as an advice by using (4) the density values and (5) specific layer combinations (Section~\ref{sec:inference}). We now detail each step in our approach.

\subsection{KDE of the Training Set} \label{sec:kde}

Given a trained classifier $M$ with $L$ layers (except for the input layer) and $C$ classes, let $\mathcal{X}^{t} = \{\mathbf{x_{1}}, \ldots, \mathbf{x_{n}}\}$ and $\mathcal{Y}^{t} = \{y_{1}, \ldots, y_{n}\}$ in $\mathit{D_{train}}$ be the set of training inputs and corresponding ground truth labels. Similarly, let $\mathcal{X}^{v}$, $\mathcal{Y}^{v}$, and $\hat{\mathcal{Y}}^{v}$ in $\mathit{D_{valid}}$ be the validation inputs, corresponding ground truth labels, and model predictions. 

We denote the outputs of all layers in the training set as feature vectors $\mathcal{V}^{t} = \{\mathbf{v}^{t}_{1}, \ldots, \mathbf{v}^{t}_{L}\}$, where the feature vectors of the layer $l$ with $n_{l}$ neurons are $\mathbf{v}^{t}_{l} \in \mathbf{R}^{n_{l}}$. We note that the feature vectors are trivially available after each execution of the trained model $M$ over a given input. In general, $M$ focuses on different features in different layers for different classes.
SelfChecker's aim is to compute the density probability of feature vectors in each layer for each class based on the training set $\mathit{D_{train}}$. Using these density probabilities SelfChecker will then estimate how close the features in a specific layer (for a certain input) are to those of the training set.

KDE is a non-parametric method for estimating a probability density function by using a finite number of samples from a population~\cite{davis2011remarks, terrell1992variable}. 
The resulting density function allows the estimation of relative likelihood of a given random variable. In this paper we use the Gaussian kernel, which works well for the multivariate data common to most datasets and produces smooth functions.
Given a data sample $\{x_{1}, x_{2}, \ldots, x_{m}\}$, SelfChecker estimates the kernel density function $f$ as follows:

\begin{equation}
\hat{f}(x) = \frac{1}{mh}\sum_{i=1}^{m}K(\frac{x-x_{i}}{h})
\end{equation}
where $K$ is the Gaussian kernel function and $h$ is \emph{bandwidth}.

To see how a KDE with Gaussian kernels works, consider Fig.~\ref{fig:kde}. 
First, each observation in the sample is replaced with a Gaussian curve centered at that value (green curves); these work as a kernel. The green curves are then summed to compute the value of the density at each point. Fig.~\ref{fig:kde}(b) also shows the normalized curve (in blue) whose area under the curve is 1. The bandwidth parameter $h$ of the KDE controls how tightly the estimate is fit to the sample data. It corresponds to the width of the kernels (green lines in Fig.~\ref{fig:kde}(b)). Fig.~\ref{fig:kde}(c) shows that if $h$ is large, the curve is smooth but flat. And, if $h$ is small, the curve is peaked and oscillating. The choice of $h$ is based on the number of sample points and their dimensions.

For each combination of class and layer, SelfChecker uses Gaussian KDE to estimate the density function that the training data for the class induces on the layer's feature vector. Then given a test instance, SelfChecker estimates the probability density for each class within each layer from the computed density functions. Finally, SelfChecker uses these probability densities to infer classes for each layer, defined as follows:

\begin{definition}[Inferred class for a layer]\label{def:infererred_class}
	Given a test instance, the \emph{inferred class for layer $\mathit{l}$} is the class for which the test instance induces the maximum estimated probability density among $\mathit{l}$'s per-class density functions.
\end{definition}

Algorithm~\ref{alg:kde} details SelfChecker's procedure for KDE estimation and inference. Lines 1-10 show the Gaussian KDE
\begin{algorithm}
	\algsetup{linenosize=\small} \scriptsize
	
	\caption{KDE Estimation and Inference}
	
	\label{alg:kde}
	\KwIn{Input instances in $\mathit{D_{train}}$, $\mathit{D_{valid}}$: $\mathcal{X}^{t}$, $\mathcal{X}^{v}$, true labels in $\mathit{D_{train}}$: $\mathcal{Y}^{t}$\;
		Trained model $M$ with $L$ layers and $C$ classes\;
		Variance threshold: $\mathit{t_{var}}$}
	\KwOut{KDE functions for each combination of class and layer: $\mathit{kdes}$\;
		Inferred classes for all layers on $\mathit{D_{valid}}$: $\mathit{kdeInferL^{v}}$}
	\# Estimation\\
		\For{c in C}
		{
			Obtain instances $\mathcal{X}^{t}_{c}$ whose true label is c\;
			\For{l in L}
			{
				$\mathbf{v}^{t}_{lc} = M.output_{l}(\mathcal{X}^{t}_{c})$\;
				Remove elements in $\mathbf{v}^{t}_{lc}$ whose variance is less than $\mathit{t_{var}}$\; 
				$\hat{f}(x) = \frac{1}{|\mathbf{v}^{t}_{lc}|h}\sum_{i=1}^{|\mathbf{v}^{t}_{lc}|}K(\frac{x-\mathbf{v}^{t}_{lc}[i]}{h})$\;
				$\mathit{kdes[l][c]} = \hat{f}(x)$\;
			}
		}

	\# Inference\\
	\For{$\mathbf{x}$ in $\mathcal{X}^{v}$}
	{
		\For{l in L}
		{
			$\mathbf{v}_{l} = M.output_{l}(\mathbf{x})$\;
			Remove values of the neurons filtered in the training set from $\mathbf{v}_{l}$\;
			\For{c in C}
			{
				$\mathit{kde\_values[c] = kdes[l][c](\mathbf{v}_{l})}$\;
			}
			$\mathit{kdeInferL^{v}[\mathbf{x}.index][l] = max(kde\_values).index}$\;
		}
	}
	
\end{algorithm}
used to extrapolate the density distribution functions of feature vectors per class in each layer. As illustrated with Fig.~\ref{fig:motivate}, we want to extrapolate the patterns of the attention overlaid on the raw input. Since the input instances with different classes perform differently in different layers, the attentions in the first layer of digit 3 are different from the first one of 6 that is also different from the second one of 6 itself. SelfChecker therefore splits the original training input instances according to their true classes (Line 3). Based on these it obtains the outputs of each layer given the trained model $\mathit{M}$ (Line 5). SelfChecker also uses mean-pooling to reduce dimensions for convolutional layers and then filters out neurons whose values show variance lower than a pre-defined threshold, $t_{var}$, to reduce the dimension of feature vectors as these neurons do not contribute much information to the KDE (Lines 6). SelfChecker then uses the filtered feature vectors to extrapolate the density functions for each layer and class, and stores them (Lines 7-8) so that they can be used for inference on new examples, such as $\mathit{D_{valid}}$ (Lines 11-21).

During inference on a given input instance, SelfChecker first obtains the outputs in each layer (Line 14), from which it removes the values of the neurons filtered in Line 6 (Line 15). It then generates the estimated density values of each class, given the corresponding KDE functions (Lines 16-18). Finally, the layer inference for the input instance is the class that has the maximum density value (Line 19), which indicates that the feature vectors of the input instance in this layer are close to those in training set that belong to this specific class. For instance, in Fig.~\ref{fig:motivate}, the class inferences given by Algorithm~\ref{alg:kde} in the three layers are 3, 3, 2 for digit 3, and 1, 6, 6 for digit 6, respectively.

\subsection{Layer Selection} \label{sec:search}

In Section~\ref{sec:motivation} we noted that different layers have different attentions, but some of these focus on a particular part of the image and may be misleading. For example, in Fig.~\ref{fig:motivate} the second and third layers for 6 are different from the final prediction. If SelfChecker would consider the outputs of these layers, it can detect that the model is not confident about the final output. And, if SelfChecker considers just these layers and uses maximum voting, then it can also provide an alternative prediction that correctly classifies this image. Therefore, the design of \emph{robust layer selection} in SelfChecker is important to accurately raise an alarm and to provide a high quality advice.


We first explain what we mean by a model output's \emph{confidence}.
Our definition is based on an observation: given a test instance, if the features of DNN layers are different from the final prediction, then the decision made by the model on the test instance will tend to be incorrect.
For example, in Fig.~\ref{fig:motivate} the attentions in the second
and third images of a 6 are more similar to those of a 6 instead of
the final prediction of 1. In this case the model misclassifies the 6
as 1.
We evaluated this observation by using Spearman rank-order correlation
coefficient and p-values~\cite{zwillinger1999crc}. Spearman rank-order measures the
relationship between the prediction correctness and the consistency of
inferred layer classes and final predictions.
Our results show that they are correlated with p-value much less than
0.05 (at most 3.09e-26) on all evaluated four image datasets and three DNN models listed in Table~\ref{tab:dataset}.

We formally define the confidence ($\delta$) of a model output ($\hat{y}$) given a test instance $\mathbf{x}$ as follows:
%

\begin{equation}
\delta=\frac{N_{kdeInferL_{x} == \hat{y}}}{N_{selectedLayerC_{alarm}[\hat{y}]}}
\end{equation}
where $N_{kdeInferL_{x} == \hat{y}}$ is the number of selected layers whose inferred class is the same as the final prediction $\hat{y}$ and $N_{selectedLayerC_{alarm}[\hat{y}]}$ is the number of selected layers for the class $\hat{y}$. Based on the maximum voting, if $\delta$ is lower than 0.5, we say that a DNN has \emph{low confidence} in prediction $\hat{y}$ for a test instance $\mathbf{x}$.

We now discuss how SelfChecker selects the proper layer combinations for each class to reach a high alarm accuracy (Algorithm~\ref{alg:alarm}). We use the training set to estimate the density function, from which the inferred class for each layer can be obtained for a given input instance. As mentioned in Section~\ref{sec:motivation}, different layers have different attentions but some of these may be misleading, we thus use the validation dataset to select layers.
\begin{algorithm}
	\algsetup{linenosize=\small} \scriptsize
	\caption{Layer Selection for Alarm}
	
	\label{alg:alarm}
	\KwIn{Input instances in $\mathit{D_{valid}}$: $\mathcal{X}^{v}$, true labels and predictions: $\mathcal{Y}^{v}$, $\hat{\mathcal{Y}}^{v}$\;
		Total classes: $C$\; 
		Inferred classes for all layers on $\mathit{D_{valid}}$: $\mathit{kdeInferL^{v}}$}
	\KwOut{Selected layers for all classes: $\mathit{selectedLayerC_{alarm}}$}
	\For{c in C}
	{
		Obtain the indexes $\mathit{idx_{c}}$ of instances $\mathcal{X}^{v}_{c}$ whose prediction $\hat{\mathcal{Y}}^{v}$ is c\;
		Generate all kinds of layer combinations $\mathit{combL}$\;
		\For{layers $l_{s}$ in $\mathit{combL}$}
		{
			\For{l in $l_{s}$}
			{
				$\mathit{y_{s}}$.add($\mathit{kdeInferL^{v}[idx_{c}][l])}$\;
			}
			$\mathit{KdePredPos}$.add(index of $sum(y_{s}$ != $\hat{\mathcal{Y}}^{v}[idx_{c}]) >= sum(y_{s} == \hat{\mathcal{Y}}^{v}[idx_{c}]))$\;
			$\mathit{TrueMisBehavior}$.add(index of $\hat{\mathcal{Y}}^{v}[idx_{c}] != c)$\;
			$\mathit{TP}$ = $\mathit{TrueMisBehavior}$ \& $\mathit{KdePredPos}$\;
			$\mathit{FP}$ = $\lnot$$\mathit{TrueMisBehavior}$ \& $\mathit{KdePredPos}$\;
			$\mathit{FN}$ = $\mathit{TrueMisBehavior}$ \& $ \lnot $$\mathit{KdePredPos}$\;
			$\mathit{F1 = 2 * TP / (2 * TP + FN + FP)}$\;
			\If {F1 is max}
			{$\mathit{selectedLayerC_{alarm}[c] = l_{s}}$\;}
		}
	}
\end{algorithm}
Given the validation dataset $\mathit{D_{valid}}$, SelfChecker splits the input instances into $C$ subsets based on their predictions (Line 2). SelfChecker then generates all possible layer combinations with lengths in range 1 through $L$, from which it searches for the best combination for each class to reach the highest accuracy (Lines 4-17). To calculate the alarm accuracy, SelfChecker first obtains the inferred class of each layer in the given layer combination (Lines 5-7) based on the generated KDE inferences across all layers on $\mathit{D_{valid}}$ ($\mathit{kdeInferL^{v}}$) by Algorithm~\ref{alg:kde}. 
To conclude whether or not the model has made a wrong prediction for an input, SelfChecker considers the layers in the layer combination. If a majority of the layers indicate inferred classes that are different from the model prediction (the confidence $\delta$ is less than 0.5), then SelfChecker concludes that the model is wrong (Line 8).
In this case, if the model prediction is indeed different from the true label of this input, the alarm is correct (True Positive), otherwise, it is incorrect (False Positive). SelfChecker uses the F1-score to measure the alarm accuracy (Lines 10-13), and it selects the layer combination with the highest accuracy for the corresponding class (Lines 14-16).

\begin{algorithm}
	\algsetup{linenosize=\small} \scriptsize
	\caption{Layer Selection for Advice}
	
	\label{alg:accuracy}
	\KwIn{Input instances in $\mathit{D_{valid}}$: $\mathcal{X}^{v}$, true labels and predictions: $\mathcal{Y}^{v}$, $\hat{\mathcal{Y}}^{v}$\;
		Total classes: $C$\; 
		Inferred classes for all layers on $\mathit{D_{valid}}$: $\mathit{kdeInferL^{v}}$\;
	Selected layers for all classes: $\mathit{selectedLayerC_{alarm}}$}
	\KwOut{Selected layers and weights per class: $\mathit{selectedLayerPosC_{advice}}$, $\mathbf{W}_{\mathit{pos}}$, $\mathit{selectedLayerNegC_{advice}}$, $\mathbf{W}_{\mathit{neg}}$\;
	}
	\For{$c_{p}$ in C}
	{
		Obtain the indexes $\mathit{idx_{c_{p}}}$ of instances $\mathcal{X}^{v}_{c_{p}}$ whose prediction $\hat{\mathcal{Y}}^{v}$ is $c_{p}$\;
	Generate $y_{s}$ given $\mathit{selectedLayerC_{alarm}[c_{p}]}$\;
		Generate all kinds of layer combinations $\mathit{combL}$\;
		$\mathit{KdePredPos}$.add(index of $sum(y_{s}$ != $\hat{\mathcal{Y}}^{v}[\mathit{idx_{c_{p}}}]) >= sum(y_{s} == \hat{\mathcal{Y}}^{v}[\mathit{idx_{c_{p}}}]))$\;
		$\mathit{TrueMisBehavior}$.add(index of $\mathcal{Y}^{v}[\mathit{idx_{c_{p}}}] != c_{p})$\;
		$\mathit{FP}$ = $ \lnot $$\mathit{TrueMisBehavior}$ \& $\mathit{KdePredPos}$\;
		\For{$c_{t}$ in C}
		{
			$\mathit{idx_{c_{t}}}$.add(index of $\mathit{KdePredPos}$ where $\mathcal{Y}^{v}[\mathit{KdePredPos}] = c_{t}$)\;
			Select layers $\mathit{selectedLayerPosC_{advice}}$ with highest accuracy $\mathit{acc_{max}}$ from $\mathit{combL}$\;
		\eIf{$c_{t} = c_{p}$}{$\mathbf{W}_{\mathit{pos}}[c_{p}][c_{t}] = len(\mathit{idx_{c_{t}}}) * \mathit{acc_{max}} / len(\mathit{KdePredPos})$}
		{$\mathbf{W}_{\mathit{pos}}[c_{p}][c_{t}] = len(\mathit{idx_{c_{t}}}) * \mathit{acc_{max}} / (len(\mathit{KdePredPos}) - \mathit{FP})$}
		}
		$\mathit{KdePredNeg}$.add(index of $sum(y_{s}$ != $\hat{\mathcal{Y}}^{v}[\mathit{idx_{c_{p}}}]) < sum(y_{s} == \hat{\mathcal{Y}}^{v}[\mathit{idx_{c_{p}}}]))$\;
		$\mathit{TN}$ = $ \lnot $$\mathit{TrueMisBehavior}$ \& $\mathit{KdePredNeg}$\;
		Iterate Lines 8-16 to obtain $\mathit{selectedLayerNegC_{advice}}$ and $\mathbf{W}_{\mathit{neg}}$
	}
	
\end{algorithm}

After selecting the layer combinations for the alarm, SelfChecker must determine the layer combinations that give a good advice whenever SelfChecker raises an alarm about a prediction. 
Algorithm~\ref{alg:accuracy} details SelfChecker's procedures for layer selection to achieve the best advice accuracy. First, SelfChecker splits the validation set $\mathit{D_{valid}}$ into $C$ subsets (Line 2), and for each subset it searches for the best layer combination. Given the layers selected for alarms by Algorithm~\ref{alg:alarm}, SelfChecker generates the KDE inferred classes in these layers as in Lines 5-7 in Algorithm~\ref{alg:alarm}. Given a test instance, if the confidence of the model prediction ($\delta$) is less than 0.5, SelfChecker concludes that the model misbehaved (Line 5). SelfChecker then searches for the best layer combination where the model predicts the input with label $c_t$ as $c_{p}$ (Lines 9-10). Since not all classes have correlation, SelfChecker obtains weights for different combinations (Lines 11-15). For example, 1 is prone to be misclassified as 7 but has little chance to be misclassified as 2.
Subsequently, in Lines 17-19, SelfChecker finds the layer combination that achieves the highest accuracy for the case where the selected layers by Algorithm~\ref{alg:alarm} indicate a negative decision (the model behaves normally). 

\textbf{Boosting strategy:} SelfChecker searches for both positive and negative decisions made by the selected layers in Algorithm~\ref{alg:alarm} in order to boost the quality of the alarm. 
In particular, if the layers selected by Algorithm~\ref{alg:alarm} indicate an alarm but the advice given by $\mathit{selectedLayerPosC_{advice}}$ (Line 10) is the same as the model prediction, then SelfChecker does not raise an alarm. Similarly, if the layers selected by Algorithm~\ref{alg:alarm} indicate that the model prediction is correct but the advice given by $\mathit{selectedLayerNegC_{advice}}$ (Line 19) is different from the model prediction, SelfChecker will raise an alarm.

\subsection{Checking the Model in Deployment} \label{sec:inference}

SelfChecker checks a trained DNN in deployment\@. It raises an alarm if it disagrees with the model's prediction of a given test instance and also generates an advice (alternative prediction). Algorithm~\ref{alg:inference} presents this process.

\begin{algorithm}
	\algsetup{linenosize=\small} \scriptsize
	\caption{Checking Model in Deployment}
	
	\label{alg:inference}
	\KwIn{Input instance and its prediction by $M$ with $L$ layers: $\mathbf{x}$, $\hat{y}$\;
		KDE functions for all layers and classes: $\mathit{kdes}$\;
		Selected layers for all classes: $\mathit{selectedLayerC_{alarm}}$, $\mathit{selectedLayerPosC_{advice}}$, $\mathit{selectedLayerNegC_{advice}}$\;
	Weights for advice: $\mathbf{W}_{\mathit{pos}}$, $\mathbf{W}_{\mathit{neg}}$}
	\KwOut{$\mathit{alarm}$ and $\mathit{advice}$ $z$}
	Generate inferred class for each layer $\mathit{kdeInferL}$ using KDE functions $\mathit{kdes}$\;
	$\mathit{L_{alarm}}$ = $\mathit{selectedLayerC_{alarm}[\hat{y}]}$\;
	Generate $y_{s}$ given $\mathit{L_{alarm}}$ and $\mathit{kdeInferL}$\;
	\eIf{$sum(y_{s}$ != $\hat{y}) >= sum(y_{s} == \hat{y}))$}
	{
		initialize $\mathit{prob}$ with $C$ dimensions\;
		\For{c in C}
		{
			$\mathit{L_{advice}}$ = $\mathit{selectedLayerPosC_{advice}}[\hat{y}][c]$\;
				\For{l in $\mathit{L_{advice}}$}
				{
					$\mathit{prob[c] = sum(kdeInferL[l] == c)}$\;
				}
			$\mathit{prob[c]} = \mathit{prob[c]} * \mathbf{W}_{\mathit{pos}}[\hat{y}][c] / len(\mathit{L_{advice}})$
		}
	$\mathit{advice = max(prob[c]).index}$\;
	\eIf{$\mathit{advice}$ != $\hat{y}$}{$\mathit{alarm = True}$, $\mathit{z = advice}$}{$\mathit{alarm = False}$}
	}
	{
		Iterate 5-18 if the alarm is not triggered initially\;
}

\end{algorithm}

First, SelfChecker generates inferred classes of all layers $\mathit{kdeInferL}$ using layer outputs and KDE functions $kdes$ obtained from Algorithm~\ref{alg:kde}. Then, as in Lines 5-7 in Algorithm~\ref{alg:alarm}, SelfChecker generates $y_{s}$ consisting of inferred classes given the selected layers for $\hat{y}$.
If the output class $\hat{y}$ is \emph{not} inferred in the majority of cases in $y_{s}$, then SelfChecker has an initial alarm that still needs to go through the boosting strategy (mentioned in the last section).

Lines 5-18 show that SelfChecker first generates the probabilities of each class given $\mathit{selectedLayerPosC_{advice}[\hat{y}]}$, which are weighted by $\mathbf{W}_{\mathit{pos}}$. If the class with the largest probability is still different from the model prediction $\hat{y}$, SelfChecker triggers the alarm and it selects the class with the largest probability as the advice. Otherwise, SelfChecker does not trigger the alarm. A similar strategy is used if the alarm is not triggered initially where the output class $\hat{y}$ is inferred in the majority of cases in $y_{s}$.

\section{Evaluation}\label{sec:exp}

In this section we present experimental evidence for the effectiveness of
SelfChecker. The goal of our evaluation is to answer the following research questions.

\begin{table*}[t]
	\small
	\caption{DL models and datasets used in the experiments.
}		
	\label{tab:dataset}	
	\centering{}%
	\begin{tabular}{p{1.8cm}<{\centering}|p{0.9cm}<{\centering}p{0.9cm}<{\centering}p{0.9cm}<{\centering}p{0.9cm}<{\centering}|p{1.2cm}<{\centering}c|p{1.2cm}<{\centering}c|p{1.2cm}<{\centering}c}
		\hline
		\multirow{3}{*}{\textbf{Dataset}} & \multirow{3}{*}{\textbf{\# Class}} & \multirow{3}{*}{\textbf{\# Train}} & \multirow{3}{*}{\textbf{\# Valid}}  & \multirow{3}{*}{\textbf{\# Test}}  & \multicolumn{6}{c}{\textbf{DL models}} \tabularnewline
		\cline{6-11}
		&  &  &  & & \multicolumn{2}{c|}{\textbf{ConvNet}} & \multicolumn{2}{c|}{\textbf{VGG-16}} & \multicolumn{2}{c}{\textbf{ResNet-20}}\tabularnewline 
		\cline{6-11}
		&  &  &  &  & \emph{\# Layers} & \emph{Accuracy\%} & \emph{\# Layers} & \emph{Accuracy\%} & \emph{\# Layers} & \emph{Accuracy\%} \tabularnewline
		\hline 
		\textbf{MNIST} & 10  & 50,000 & 10,000 & 10,000 & 8 & 99.36 & 16 & 98.87 & - & -\tabularnewline
		\hline 
		\textbf{FMNIST} & 10 & 50,000  & 10,000 & 10,000 & 8 & 92.13 & 16 & 93.75 & 20 & 92.74 \tabularnewline
		\hline 
		\textbf{CIFAR-10} & 10 & 40,000  & 10,000 & 10,000 & 8 & 80.45 & 16 & 92.17 & 20 & 92.08 \tabularnewline
		\hline
		\textbf{CIFAR-100} & 100 & 40,000  & 10,000 & 10,000 & - & - & 16 & 66.79 & 20 & 69.52 \tabularnewline
		\hline 
	\end{tabular}

\footnotesize{\vspace{0.5em}ResNet-20 and ConvNet are seldom used for MNIST and CIFAR-100. We omit their results due to space limitation but we will release them with our code. DAVE-2 and Chauffeur for self-driving cars are regression models so we exclude them in this table.}
\end{table*}

\subsection{Research Questions}

\noindent \textbf{RQ1. Alarm Accuracy:} \emph{How effective is SelfChecker in predicting DNN misclassifications in deployment?}

To evaluate the effectiveness of SelfChecker for raising alarms in deployment, we compare its alarm accuracy on the test dataset with related techniques, namely, \textsc{SelfOracle}~\cite{stocco2020misbehaviour}, \textsc{Dissector}~\cite{wang2020dissector}, and ConfidNet~\cite{corbiere2019addressing}. For the comparison, we chose the variant from \textsc{SelfOracle}---the VAE (variational autoencoder)---that achieved the best performance against other \textsc{SelfOracle} variants, with confidence threshold of 0.05. Since \textsc{Dissector} did not provide the threshold for distinguishing beyond-inputs from within-inputs, we used the validation dataset to choose a threshold in the $0 - 1$ range with the highest F1-score and the best weight growth type from \textit{linear}, \textit{logarithmic}, and \textit{exponential} defined in~\cite{wang2020dissector} with the highest Area Under Curve (AUC) for each dataset and DNN classifier. We also used the validation dataset to find the best threshold of failure prediction for ConfidNet to reach the highest F1-score.

\vspace{1em}
\noindent \textbf{RQ2. Advice Accuracy:} \emph{Does the advice given by SelfChecker improve the accuracy of a DNN?}

In cases where SelfChecker raises an alarm about a model prediction, we also determine whether it can provide an advice and the accuracy of this advice. To answer this question, we compare the advice accuracy of SelfChecker against the accuracy of the original DL model $M$\@. For self-driving cars, we use the dataset released by \textsc{SelfOracle}. This dataset only includes anomalous/normal labels, which is not enough to provide realistic advice\xy{, such as turning right/left}.

\vspace{1em}
\noindent \textbf{RQ3. Deployment Time:} \emph{What is the time overhead of SelfChecker in deployment for a given test instance?}

We consider what different algorithms do in deployment and evaluate
the computation time of their deployment-time components. SelfChecker
performs DNN computation, KDE inferences, and alarm and advice
analysis. \textsc{SelfOracle} uses the reconstructor to compute a loss
and anomaly detector. \textsc{Dissector} generates probability vectors
and performs validity analysis\footnote{By contrast, Wang et
  al.~\cite{wang2020dissector} only include validity analysis. We
  believe that the probability vector generation must also be
  performed during deployment, since this is the input to validity
  analysis.}. ConfidNet computes an output using two DNNs.

\vspace{1em}
\noindent \textbf{RQ4. Layer Selection:} \emph{Does the choice of layers for selection by SelfChecker have
  an impact on its alarm accuracy?}

Kaya et al.~\cite{kaya2019shallow} characterized "over-thinking" as a prevalent weakness of DL models, which occurs when a DL model can reach correct predictions before its final layer. Over-thinking can be destructive when a correct prediction within hidden layers changes to a misclassification at the output layer (see Section~\ref{sec:motivation}). Therefore, it is important to select proper layers for different classes. To evaluate the impact of layer selections on the alarm accuracy, we experimented
with three layer selection strategies \xy{as discussed in Section \ref{sec:RA}: RQ4}. 

\vspace{1em}
\noindent \textbf{RQ5. Boosting Strategy:} \emph{Does the boosting strategy improve SelfChecker's alarm accuracy, particularly in terms of decreasing the number of false alarms?}

As discussed in Sections~\ref{sec:search} and~\ref{sec:inference}, we use a boosting strategy to check whether or not to raise an alarm.

\subsection{Experimental Setup}\label{sec:setup}

We evaluate SelfChecker on four popular datasets (MNIST~\cite{lecun2010mnist}, FMNIST~\cite{xiao2017fashion}, CIFAR-10~\cite{krizhevsky2009learning}, and CIFAR-100~\cite{krizhevsky2009learning}) using three DL models (ConvNet~\cite{kim2019guiding}, VGG-16~\cite{simonyan2014very}, and ResNet-20~\cite{he2016deep}). We also compare the alarm accuracy of SelfChecker against \textsc{SelfOracle}~\cite{stocco2020misbehaviour} for self-driving car scenarios evaluated on two publicly-available DL models, NVIDIA's DAVE-2~\cite{bojarski2016end} and Chauffeur~\cite{chauffeur/online}.  To reduce the possibility of fluctuation due to randomness, we ran all experiments involving MNIST, FMNIST, CIFAR-10, and CIFAR-100 three times and computed the average of all metrics.  For the experiments involving the driving datasets, we ran each experiment just once, since we used pre-trained models released by the authors of \textsc{SelfOracle}~\cite{stocco2020misbehaviour}.
We conducted all experiments on an Ubuntu 18.04 server with Intel i9-10900X (10-core) CPU @ 3.70GHz, one RTX 2070 SUPER GPU, and 64GB RAM.

\begin{table*}[htb]
	\small
	\caption{Alarm accuracy.}	
	\label{tab:alarm}	
	\centering{}%
	\begin{tabular}{|p{1.5cm}<{\centering}|p{1.5cm}<{\centering}|cccc|cccc|cccc|}
		\hline 
		\multirow{2}{*}{\textbf{Dataset}}& \multirow{2}{*}{\textbf{DL}}&\multicolumn{4}{c|}{\textbf{$\uparrow$ TPR \%}}&\multicolumn{4}{c|}{\textbf{$\downarrow$ FPR \%}}&\multicolumn{4}{c|}{\textbf{$\uparrow$ F1 \%}}\tabularnewline
		\cline{3-14} 
		&  & \textit{SO} & \textit{DT} & \textit{CN} & \textit{SC} & \textit{SO} & \textit{DT} & \textit{CN} & \textit{SC} & \textit{SO} & \textit{DT} & \textit{CN} & \textit{SC} \tabularnewline
		\hline 
		\multirow{2}{*}{\textbf{MNIST}} & \textit{ConvNet} & 18.75 & 60.94 & 60.94 & \textbf{62.50 }& 4.39 & 0.24 & 0.58 & \textbf{0.23} & 4.69 & 61.42 & 48.45 & \textbf{62.99} \tabularnewline
		\cline{2-14} 
		& \textit{VGG-16} & 20.35 & 68.14 & 61.95 & \textbf{74.34} & 4.29 & 0.32 & 0.46 & \textbf{0.31} & 8.21 & 69.37 & 61.40 & \textbf{73.68} \tabularnewline
		\hline 
		\multirow{3}{*}{\textbf{FMNIST}} & \textit{ConvNet} & 9.53  & \textbf{47.65} & 38.12 & 41.55 & 5.60 & 4.03 & 0.73 & \textbf{0.51} & 10.89 & 48.92 & 51.99 & \textbf{56.33}\tabularnewline
		\cline{2-14} 
		& \textit{VGG-16} & 8.00 & \textbf{48.48} & 43.36 & 46.88 & 5.75 & 4.16 & 0.98 & \textbf{0.86} & 8.24 & 45.98 & 54.86 & \textbf{58.66} \tabularnewline
		\cline{2-14}  
		& \textit{ResNet-20} & 9.64 & \textbf{54.96} & 47.66 & 51.79 & 5.69 & 3.76 & 1.14 & \textbf{0.98} & 10.57 & 54.14 & 58.74 &\textbf{63.03} \tabularnewline
		\hline 
		\multirow{3}{*}{\textbf{CIFAR-10}} & \textit{ConvNet} & 5.01 & 61.43 & 58.57 & \textbf{61.89} & 3.97 & 9.83 & 2.29 & \textbf{2.04 }& 8.26 & 60.86 & 69.73 & \textbf{72.69}\tabularnewline
		\cline{2-14} 
		& \textit{VGG-16} & 6.39 & \textbf{53.77} & 43.17 & 49.30 & 3.94 & 4.47 & 3.03 & \textbf{1.16} & 8.36 & 52.10 & 48.29 & \textbf{60.50} \tabularnewline
		\cline{2-14} 
		& \textit{ResNet-20} & 7.07 & 47.98 & 49.87 & \textbf{52.15} & 3.96 & 4.93 & 1.03 & \textbf{0.64} & 9.23 & 46.74 & 61.62 & \textbf{65.35} \tabularnewline
		\hline 
		\multirow{2}{*}{\textbf{CIFAR-100}} & \textit{VGG-16} & 10.48 & 82.78 & 78.20 & \textbf{84.22} & 7.88 & 23.78 & 16.17 & \textbf{6.57} & 16.59 & 71.79 & 74.22 & \textbf{85.31}\tabularnewline
		\cline{2-14} 
		& \textit{ResNet-20} & 11.25 & 75.16 & 61.15 & \textbf{80.97} & 7.64 & 21.63 & 13.56 & \textbf{7.09} & 17.49 & 66.96 & 63.67 & \textbf{82.14} \tabularnewline
		\hline 
		\multirow{2}{*}{\textbf{Driving}} & \textit{DAVE-2} & 76.85  & - & - & \textbf{99.01} & \textbf{7.29}  & - &  - & 9.37 & 46.43 & - & - &  \textbf{49.88} \tabularnewline
		\cline{2-14} 
		& \textit{Chauffeur} & 81.15 & - & - & \textbf{93.44} & 4.77 & - & - & \textbf{4.56} & 32.25 & - & - & \textbf{37.25}  \tabularnewline
		\hline 
	\end{tabular}
	
	\footnotesize{\vspace{0.5em}SO, DT, CN, and SC stand for \textsc{SelfOracle}, \textsc{Dissector}, ConfidNet, and SelfChecker, respectively.}
\end{table*}

\vspace{.2em}
\noindent \textbf{Datasets and DL models.}
Table~\ref{tab:dataset} lists the number of classes and the number of training, validation, and test instances in each dataset, as well as the number of layers and the testing accuracy of all trained DL models. These datasets are widely used and each is a collection of images. ConvNet, VGG-16, and ResNet-20 are commonly-used DL models whose sizes range from small to large, with the number of layers ranging from 8 to 20. Table~\ref{tab:dataset} presents the accuracy of each model we obtained for each dataset; these accuracies are similar to the state-of-the-art.
As mentioned in Section \ref{sec:design}, SelfChecker has a training module and a deployment module. The training and validation dataset were used in the training module, and the test dataset were used on the deployment module to evaluate the performance of SelfChecker.



For our experiments with NVIDIA's DAVE-2~\cite{bojarski2016end} and Chauffeur~\cite{chauffeur/online} for self-driving cars, we used the dataset and pre-trained models released by the authors of \textsc{SelfOracle}. There are 37,947 training images, 9,486 validation images and 134,820 testing images for DAVE-2 and 250,830 for Chauffeur. The testing images are collected by the self-driving car respectively equipped with the two trained DL models. The collection process stops when the car has collisions or out-of-bound episodes. Therefore, the testing images are different for the two DL models.
DAVE-2 contains five convolutional layers followed by three fully-connected layers, while Chauffeur consists of six convolutional layers followed by one fully-connected layer.

\vspace{.2em}
\noindent \textbf{Configurations.}
As discussed in Section~\ref{sec:design}, we filter out neurons whose activation values show variance lower than a pre-defined threshold ($t_{var}$ in Algorithm~\ref{alg:kde}), as these neurons do not contribute much information to the KDE\@. For all research questions, the default variance threshold is set to $10^{-5}$, and the bandwidth for KDE is set using Scott's Rule~\cite{scott2015multivariate} based on the number of data points and dimensions.

\vspace{.2em}
\noindent \textbf{Metrics.}
Given the KDE inferences of the selected layers, if more layers disagree than agree with the model output, SelfChecker triggers an alarm. We compute the confusion metrics (TP, FP, TN, and FN) as our measurement. Consequently, a True Positive (TP) is defined when SelfChecker triggers an alarm to predict a misclassification where the model output is indeed wrong. Conversely, a False Negative (FN) occurs when SelfChecker does not trigger an alarm on a real misclassification by the model. A False Positive (FP) represents a false alarm by SelfChecker, whereas True Negative (TN) cases occur when SelfChecker is silent on correct classifications.
Our goal is to achieve (1)~a high true positive rate (TPR = TP / (TP+FN)),
(2)~a low false positive rate (FPR = FP / (TN+FP)), and
(3)~a high F1-score (F1 = (2 * TP) / ((2 * TP) + FN + FP)).

\subsection{Results and Analyses}\label{sec:RA}

We now present results that answer our research questions.

\vspace*{0.2em}\noindent\textbf{RQ1. Alarm Accuracy}\vspace*{0.2em}

\begin{figure*}[t]
 	\centering
 	\subfigure[\# True Positives (TP)]{
 		\begin{minipage}[t]{0.23\linewidth}
 			\centering
 			\includegraphics[width=4.55cm]{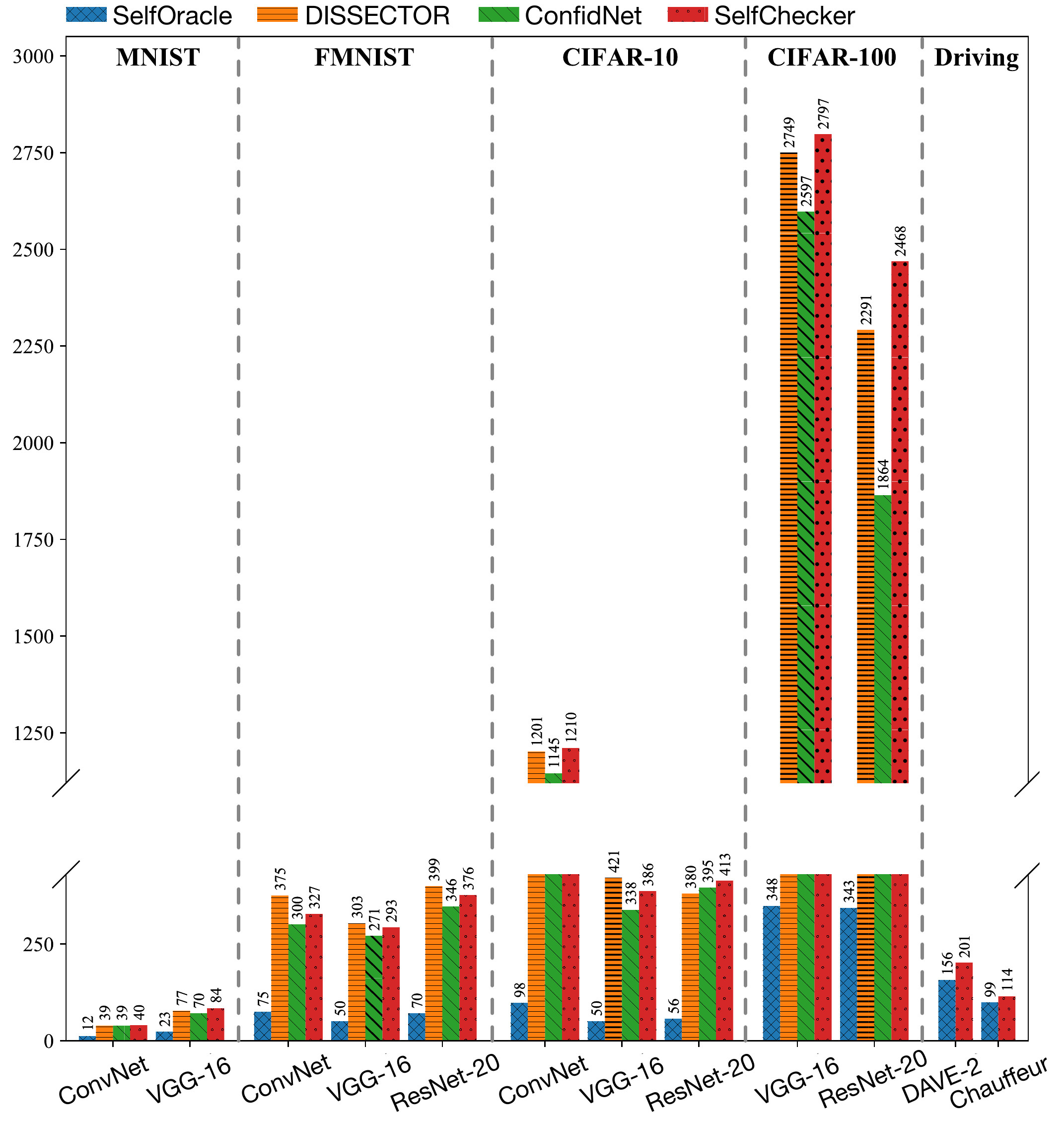} 
 		\end{minipage}
 	}
 	\subfigure[\# False Positives (FP)]{
 		\begin{minipage}[t]{0.23\linewidth}
 			\centering
 			\includegraphics[width=4.55cm]{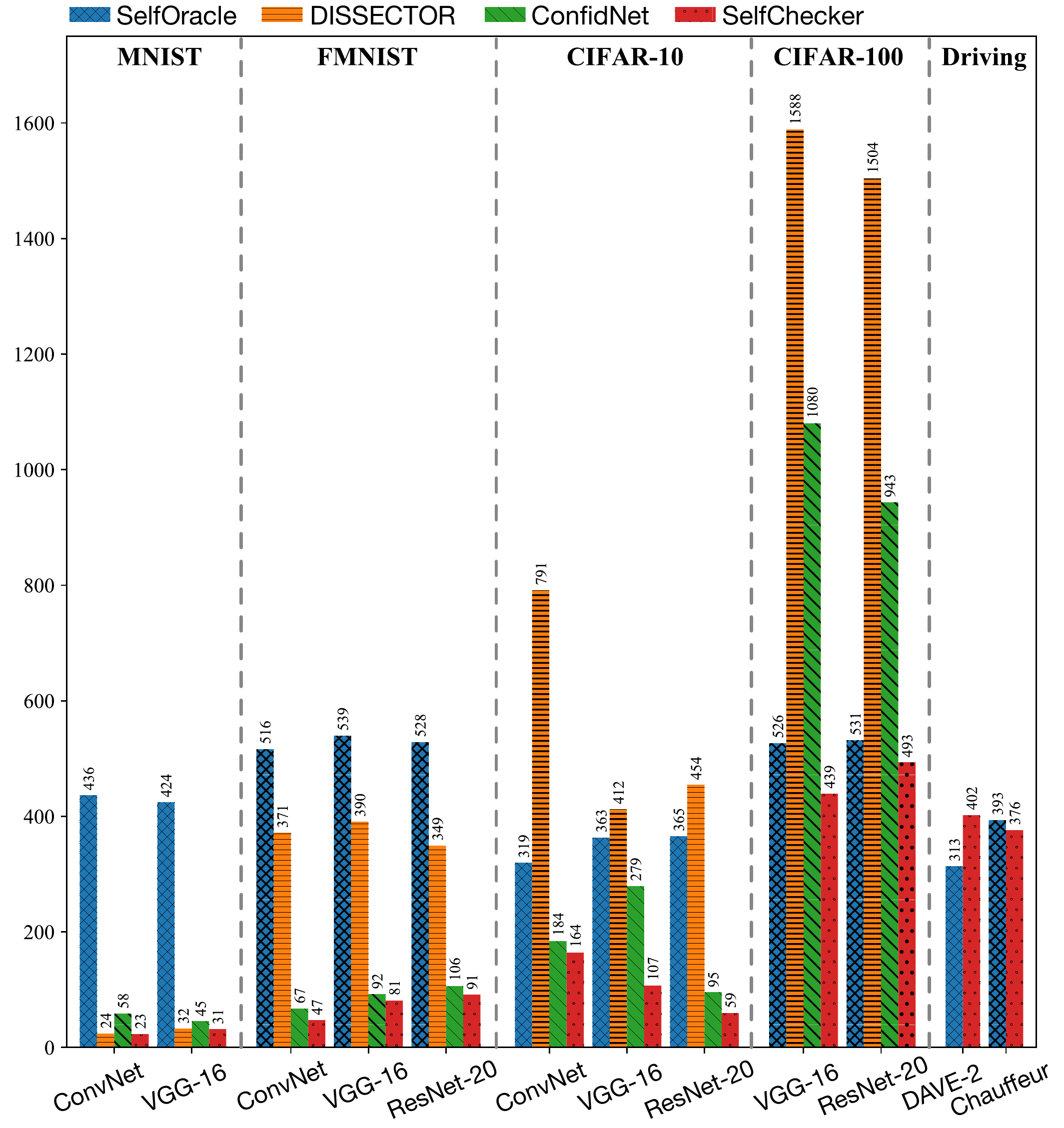} 
 		\end{minipage}
 	}
 	\subfigure[\# True Negatives (TN)]{
 		\begin{minipage}[t]{0.23\linewidth}
 			\centering
 			\includegraphics[width=4.55cm]{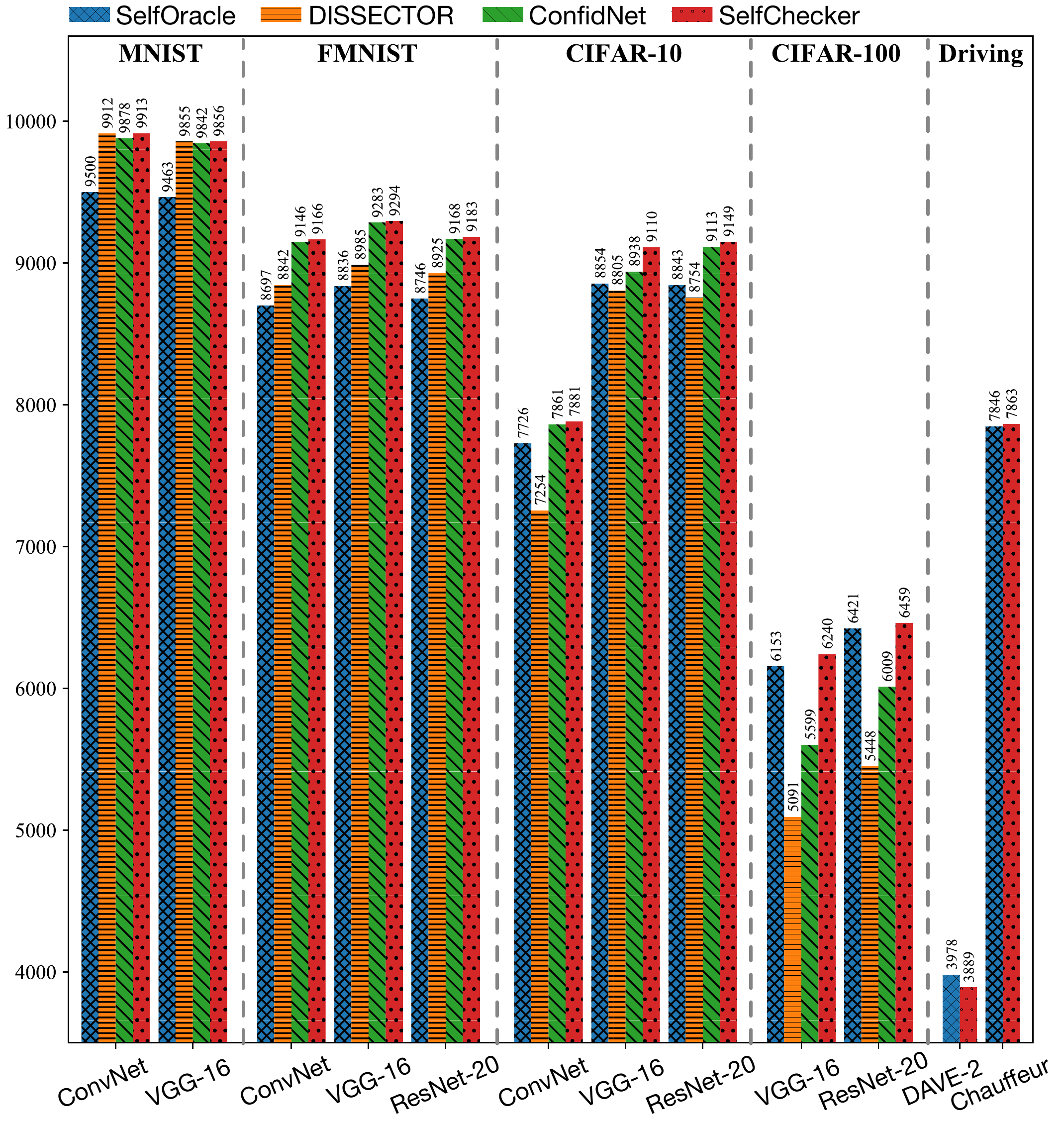} 
 		\end{minipage}
 	}
 	\subfigure[\# False Negatives (FN)]{
 		\begin{minipage}[t]{0.23\linewidth}
 			\centering
 			\includegraphics[width=4.55cm]{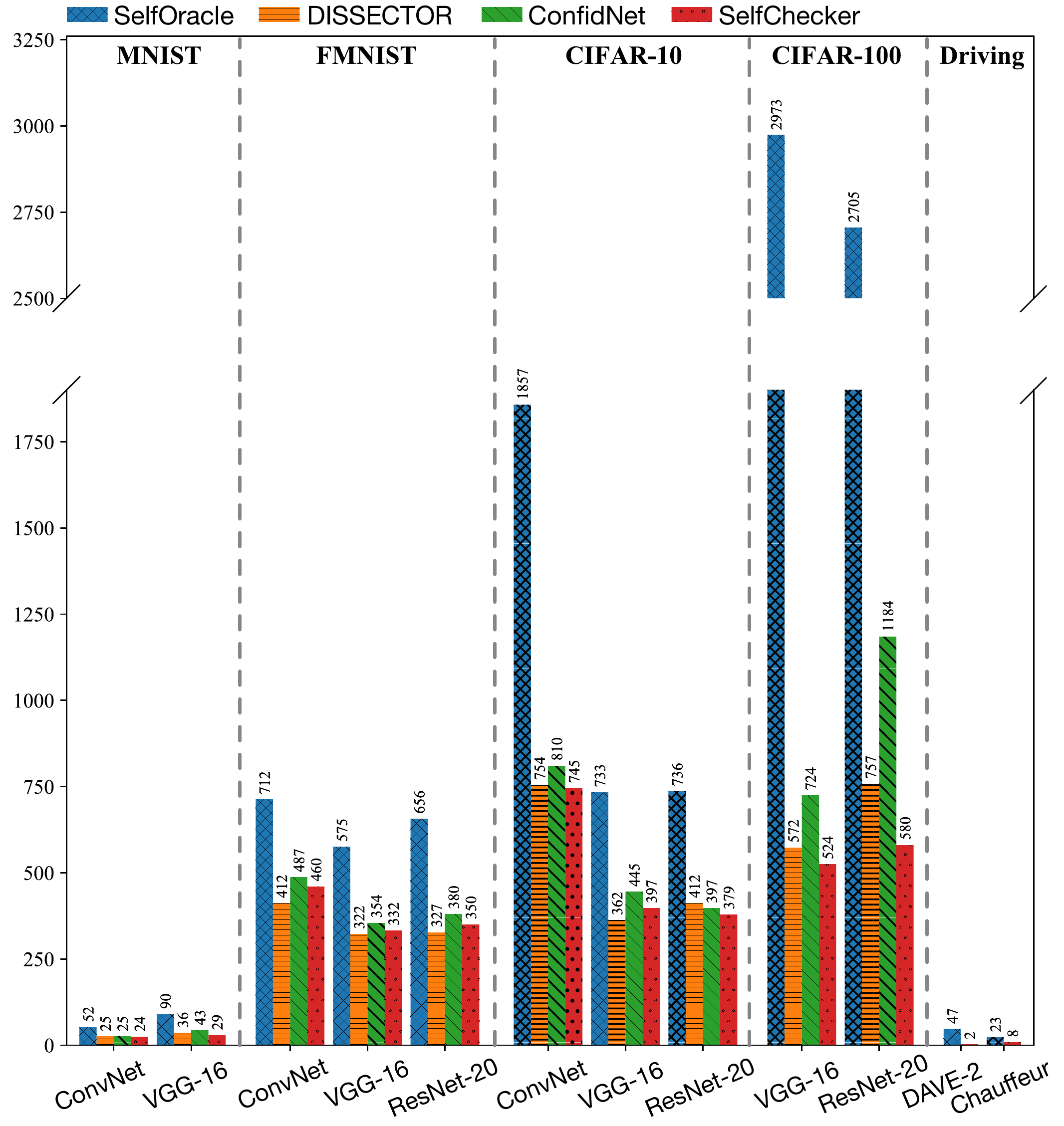} 
 		\end{minipage}
 	}
 	\caption{Confusion metrics comparing the performance of all approaches.}
 	\label{fig:alarm}
 \end{figure*}

\noindent Table~\ref{tab:alarm} presents the alarm accuracies of three DL models (ConvNet, VGG-16, and ResNet-20) in deployment on four datasets (MNIST, FMNIST, CIFAR-10, and CIFAR-100) checked by \textsc{SelfOracle}, \textsc{Dissector}, ConfidNet and SelfChecker, and the alarm accuracies of two self-driving car DL models checked by SelfChecker and \textsc{SelfOracle}~\cite{stocco2020misbehaviour}, in terms of TPR, FPR and F1-score. Fig. \ref{fig:alarm} shows the average confusion metrics of all datasets and DL models. SelfChecker can always trigger more correct alarms (TP) and miss fewer true alarms (FN) than \textsc{SelfOracle} and ConfidNet.

On traditional DNN classifiers, SelfChecker correctly triggers an alarm on over half of the misclassifications (average TPR 60.56\%), which is much higher than that of \textsc{SelfOracle} (average TPR 10.65\%) and ConfidNet (average TPR 54.30\%), and comparable to \textsc{Dissector} (average TPR 60.13\%). In particular, the highest TPR of SelfChecker is 84.22\%; this means that over 80\% of misclassifications can be detected by SelfChecker. 
However, there are four cases on which \textsc{Dissector} achieves higher TPR. Similar to SelfChecker, \textsc{Dissector} also benefits from the internal layer features. It builds several sub-models that are retrained on top of internal layers. Therefore, additional information may be learned by the training process that SelfChecker lacks. But, SelfChecker outperforms \textsc{Dissector} on TPR in the majority of cases, which indicates that the additional information is limited. Significantly, SelfChecker outperforms \textsc{SelfOracle}, which has no internal information and ConfidNet, which only considers high-level representations on all datasets and DNN classifiers on TPR. We thus conclude that the internal layer features obtained by SelfChecker are important to detecting misclassifications.
On the other hand, SelfChecker achieves lower FPR than all the competitors. 
The low FPR indicates that SelfChecker triggers few false alarms. This is expected since the boosting strategy (Section~\ref{sec:search}) makes SelfChecker very prudent in triggering alarms.
Finally, SelfChecker has a higher F1-score than all the competing approaches with an average values of 68.07\% against 10.25\%, 57.83\%, and 59.30\% for \textsc{SelfOracle}, \textsc{Dissector}, and ConfidNet, respectively.
The reason \textsc{SelfOracle} has worse accuracy on traditional DNN classifiers is that it is tailored for time series analysis on video frame sequences that change little over short periods of time. ConfidNet is trained on top of the original DL model whose weights of feature extraction are frozen using the training dataset and it uses the loss function based on true class probability. Since there are few wrong predictions in the training dataset after the original model is trained, overfitting leads to limited performance of ConfidNet. Note that the results of ConfidNet shown in Table~\ref{tab:alarm} are different from those in~\cite{corbiere2019addressing} since our study regards wrong predictions as positive cases (discussed in Metrics in Section~\ref{sec:setup}) while~\cite{corbiere2019addressing} regards correct predictions as positive cases.

In the self-driving car scenarios, we transformed the regression network that predicts steering angles into a binary classification network that classifies steering angles as either normal or anomalous. Since the true class probability is the base of ConfidNet, and the first and second highest class probabilities are necessary for \textsc{Dissector}, both of these cannot be used in the self-driving car scenarios.
Given the validation dataset, a Gamma distribution is fitted to the errors between the predictions and the real-valued angles (MSE), and density values of each layer generated by Algorithm~\ref{alg:kde}, respectively. Given an $\epsilon$ value of 0.05 (the same as used in \textsc{SelfOracle}) from the Gamma fitting distribution, if the error of an instance in the validation dataset is larger than the value corresponding to $\epsilon$, it is labeled as an anomaly. Similarly, if the density value is less than the values corresponding to $\epsilon$, it is predicted as an anomaly. 
We then use SelfChecker to solve the regression problem as a binary classification problem.
Table \ref{tab:alarm} shows that SelfChecker achieves a higher TPR than \textsc{SelfOracle} on both DAVE-2 and Chauffeur, indicating that SelfChecker can trigger more correct alarms. Even though SelfChecker triggers more false alarms for DAVE-2, it also triggers more true alarms (201 against 156 by \textsc{SelfOracle}) and misses only 2 true alarms. In addition, the F1-score for SelfChecker is higher than for \textsc{SelfOracle} on both models.

For RQ1, we conclude that SelfChecker effectively triggers alarms that predict misbehaviors of DL models in deployment with high TPR and low FPR\@.

\vspace*{0.2em}\noindent\textbf{RQ2. Advice Accuracy}\vspace*{0.2em}

\noindent Table~\ref{tab:advice} compares the accuracies of the original model $M$ to those of $M$ having advice provided by SelfChecker. Even though SelfChecker achieves high alarm accuracies, it is challenging for it to provide correct advice as we regard the advice as correct only if the inferred classes of most selected internal layers are the same as the true label. This condition is more strict than triggering an alarm that requires the inferred classes of most selected internal layers to be different from the model's prediction.
\begin{table}[h]
	\small
	\caption{Advice accuracy.}	
	\label{tab:advice}	
	\centering{}%
	\begin{tabular}{|c|c|c|c|c|}
		\hline 
		\textbf{Accuracy} & \textbf{Strategies} & \textbf{ConvNet} & \textbf{VGG-16} & \textbf{ResNet-20} \tabularnewline
		\hline 
		\multirow{2}{*}{\textbf{MNIST}} & $M$ & 99.36 & 98.87  & - \tabularnewline
		\cline{2-5} 
		& \textit{$M$+SC} & \textbf{99.37} &  \textbf{99.21} & - \tabularnewline
		\hline 
		\multirow{2}{*}{\textbf{FMNIST}} & $M$ & 92.13 & 93.75 & 92.74  \tabularnewline
		\cline{2-5} 
		& \textit{$M$+SC} & \textbf{92.34}  & \textbf{93.78} & \textbf{92.80} \tabularnewline
		\hline 
		\multirow{2}{*}{\textbf{CIFAR-10}} & $M$ & 80.45 & 92.17  & 92.08 \tabularnewline
		\cline{2-5} 
		& \textit{$M$+SC} & \textbf{80.63} & \textbf{92.41}  & \textbf{92.11} \tabularnewline
		\hline 
		\multirow{2}{*}{\textbf{CIFAR-100}} & $M$ & - & \textbf{66.79} & \textbf{69.52} \tabularnewline
		\cline{2-5} 
		& \textit{$M$+SC} & - & 66.16 & 68.85 \tabularnewline
		\hline 
	\end{tabular}
	
	\footnotesize{\vspace{0.5em}SC stands for SelfChecker.}
\end{table}
Our results show that even though the trained DL models have achieved state-of-the-art accuracies, the advice can still improve model's prediction accuracy by about 0.138\% for datasets with 10 classes but decrease the prediction accuracy by about 0.65\% for datasets with 100 classes. There are two reasons for this. First, finding a correct prediction from 100 classes is a harder problem. Second, the validation set per class is more limited: CIFAR-10 has 1000 samples per class but CIFAR-100 only has 100 samples per class. 
We empirically find that SelfChecker's advice can improve model's prediction accuracy when the number of samples per class is over 200.
The results also show that the advice provided by SelfChecker can improve the prediction accuracy at most 0.34\% without retraining with additional inputs or changing the architecture.
Even though this difference is small, for a safety-critical domain such as self-driving cars, which make tens of decisions per second, a difference of 0.2\% in 10,000 decisions translates to 20 fewer misclassifications. 

For RQ2, we showed that SelfChecker's advice can improve the accuracy of the original models beyond their state-of-the-art performance with a sufficiently large validation dataset.

\vspace*{0.2em}\noindent\textbf{RQ3. Deployment Time}\vspace*{0.2em}

We measured the average time that it takes a method to check a model's
inference on a single input. Table~\ref{tab:time} lists the average times for all the datasets in Table \ref{tab:alarm} for each DNN classifier. The results for DAVE-2 and Chauffeur are for their corresponding self-driving datasets.
%
\textsc{SelfOracle} and ConfidNet take the least time since they use
an additional DL model and their deployment checking time is the time
it takes for two DL models to compute their outputs. However, these
methods have alarm accuracies that are lower than \textsc{Dissector}
and SelfChecker. \textsc{Dissector} takes longer than SelfChecker
(average of 50.47ms vs 34.98ms) on traditional DNN classifiers. 

\begin{table}[h]
	\small
	\caption{Deployment time.}	
	\label{tab:time}	
	\centering{}%
	\begin{tabular}{|c|c|c|c|c|}
		\hline 
		\textbf{Time (ms)} & \textbf{SO} & \textbf{DT} & \textbf{CN} & \textbf{SC} \tabularnewline
		\hline 
		\textbf{ConvNet} & 0.96 & 29.74 & 0.98 & 26.47  \tabularnewline
		\hline 
		\textbf{VGG-16} & 1.35 & 58.34 & 1.02 & 35.83  \tabularnewline
		\hline 
		\textbf{ResNet-20} & 1.79 & 63.33 & 1.36 & 42.63  \tabularnewline
		\hline
		\textbf{DAVE-2} & 45.80 & - & - & 67.78  \tabularnewline
		\hline
		\textbf{Chauffeur} & 42.66 & - & - & 63.12  \tabularnewline
		\hline
	\end{tabular}
	
	\footnotesize{\vspace{0.5em}
		SO, DT, CN, and SC stand for \textsc{SelfOracle}, \\
		\textsc{Dissector}, ConfidNet, and SelfChecker, respectively}
\end{table}

We believe that these checking times are acceptable across a
variety of application domains. 
As is, SelfChecker can be used for applications ranging from medical image-based diagnosis to airport security screening. For real-time applications (e.g., autonomous driving), the latency of SelfChecker and \textsc{SelfOracle} needs to improve. The checking time in the self-driving
car scenarios is high because 32 frames must be analyzed before raising an alarm. 
Efficiency is not this paper’s focus, but we acknowledge its importance for cyber-physical systems. We plan to parallelize SelfChecker by using a process per class density function to decrease latency by 1/(number of classes).

\vspace*{0.2em}\noindent\textbf{RQ4. Layer Selection}\vspace*{0.2em}

\noindent As discussed in Section~\ref{sec:search}, we use search-based optimization to select suitable layers for improving alarm accuracy. We present the results of checking VGG-16 on FMNIST and Chauffeur on the self-driving car dataset in Table~\ref{tab:selection}; \xy{we omit results for the other models and dataset since they have similar properties.}
We evaluate three layer selection strategies for triggering alarms and compare them in terms of alarm accuracy. The first strategy involves random selection of layers for each class, with the number of layers selected for each class being the same as the number selected using our approach, in order to make a fair comparison. The second strategy uses the full set of layers. The third strategy is our own approach described in Section~\ref{sec:search}, which selects suitable layers based on the validation dataset. To ensure a fair comparison, none of the strategies use the boosting strategy.

\begin{table}[h]
	\small
	\caption{Impact of layer selection on alarm accuracy.}	
	\label{tab:selection}	
	\centering{}%
	\begin{tabular}{|p{1.4cm}<{\centering}|p{0.34cm}<{\centering}p{0.34cm}<{\centering}p{0.36cm}<{\centering}p{0.34cm}<{\centering}|c|c|c|}
		\hline 
		\textbf{FMNIST} & \textbf{TP} & \textbf{FP} & \textbf{TN} & \textbf{FN} & \textbf{$\uparrow$ TPR} & \textbf{$\downarrow$ FPR} & \textbf{$\uparrow$ F1}\tabularnewline
		\hline 
		\textbf{Random} & 280 & 482 & 8893 & 345 & 44.80 & 5.14 & 40.37 \tabularnewline
		\hline 
		\textbf{Full} & 209 & 230 & 9145 & 416 & 33.44 & \textbf{2.45} & 39.29 \tabularnewline
		\hline 
		\textbf{SC-layer}$^a$ & 317 & 329 & 9046 & 308 & \textbf{50.72} & 3.51 & \textbf{49.88}\tabularnewline
		\hline\hline 
		\textbf{Chauffeur} & \textbf{TP} & \textbf{FP} & \textbf{TN} & \textbf{FN} & \textbf{$\uparrow$ TPR} & \textbf{$\downarrow$ FPR} & \textbf{$\uparrow$ F1}\tabularnewline
		\hline 
		\textbf{Random} & 112 & 3059 & 5180 & 10 & 91.80 & 37.13 & 6.80 \tabularnewline
		\hline 
		\textbf{Full} & 99 & 2596 & 5643 & 23 & 81.15 & \textbf{31.51} & 7.03 \tabularnewline
		\hline 
		\textbf{SC-layer}$^a$ & 116 & 2978 & 5261 & 6 & \textbf{95.08} & 36.15 & \textbf{7.21}\tabularnewline
		\hline 
	\end{tabular}

\footnotesize{\vspace{0.5em}$^a$ SC-layer stands for SelfChecker's layer selection. 
}
\end{table}

The results in Table~\ref{tab:selection} indicate that SelfChecker's layer selection strategy always achieves the highest TPR and F1-score compared to random selection and full selection. Even though using all layers to decide whether triggering an alarm achieves lower FPR than our approach, it sacrifices the number of correct alarms by 108 and 17 for FMNIST and driving dataset, respectively. Therefore, selecting more layers does not lead to a better checker.

For RQ4, we conclude that a careful selection of layers allows SelfChecker to identify more misclassifications and raise more correct alarms.

\vspace*{0.2em}\noindent\textbf{RQ5. Boosting Strategy}\vspace*{0.2em}

\noindent Table~\ref{tab:boosting} presents the alarm accuracies of SelfChecker both with (\emph{SC}) and without (\emph{SC\textit{-b}}) the boosting strategy  described in Section~\ref{sec:search}, for ResNet-20 on FMNIST and CIFAR-100; \xy{we omit results for the other models and dataset since they have similar properties.} As indicated in Table~\ref{tab:boosting}, adopting the boosting strategy achieves much lower FPR \xy{(the lower the better)} than \emph{SC\textit{-b}}, with larger F1-score \xy{(the higher the better)}\@. 

\begin{table}[h]
	\small
	\caption{Impact of boosting on alarm accuracy checking ResNet-20.}	
	\label{tab:boosting}	
	\centering{}%
	\begin{tabular}{|p{1.2cm}<{\centering}|p{0.34cm}<{\centering}p{0.34cm}<{\centering}p{0.36cm}<{\centering}p{0.34cm}<{\centering}|c|c|c|}
		\hline 
		\textbf{FMNIST} & \textbf{TP} & \textbf{FP} & \textbf{TN} & \textbf{FN} & \textbf{$\uparrow$ TPR} & \textbf{$\downarrow$ FPR} & \textbf{$\uparrow$ F1}\tabularnewline
		\hline 
		\textbf{SC\textit{-b}} &402 & 323 & 8951 & 324 & 55.37 & 3.48 & 55.41 \tabularnewline
		\hline 
		\textbf{SC} & 376 & 91 & 9183 & 350 & 51.79 & \textbf{0.98} & \textbf{63.03} \tabularnewline
		\hline\hline 
		\textbf{CIFAR}$^a$ & \textbf{TP} & \textbf{FP} & \textbf{TN} & \textbf{FN} & \textbf{$\uparrow$ TPR} & \textbf{$\downarrow$ FPR} & \textbf{$\uparrow$ F1}\tabularnewline
		\hline 
		\textbf{SC\textit{-b}}  & 2571 & 930 & 6022 & 477 & 84.35 & 13.38 & 78.52 \tabularnewline
		\hline 
		\textbf{SC} & 2468 & 493 & 6459 & 580 & 80.97 & \textbf{7.09} & \textbf{82.14} \tabularnewline
		\hline 
	\end{tabular}

\footnotesize{\vspace{0.5em}$^a$ CIFAR stands for CIFAR-100}
\end{table}

For RQ5, we showed that the boosting strategy significantly improves alarm accuracy by reducing false alarms.

\section{Related Work}

Most studies that check DL model trustworthiness focus on the process of model engineering: generate adversarial test instances~\cite{goodfellow2014explaining, zhang2018deeproad,moosavi2016deepfool,nguyen2015deep,wicker2018feature,li2020adversarial}, increase test coverage~\cite{tian2018deeptest, ma2018combinatorial,pei2017deepxplore}, and improve robust accuracy~\cite{ma2018mode, kim2019guiding}. Unlike our work, which checks the model in production, these approaches rely heavily on manually supplied ground truth labels.
Our focus is on non-adversarial inputs, which require different considerations~\cite{zhang2020towards}. We plan to consider adversarial inputs in our future work.
%
%

SelfChecker’s performance will depend on the difference in distribution. We conducted preliminary experiments by \textit{slightly} changing the testing dataset with random noise to push the dataset embeddings of the first fully-connected layer after all convolutional layers away from the training dataset. In this setup, SelfChecker performs similarly to the normal in-distribution dataset. Besides, there are existing studies detecting out-of-distribution data~\cite{liang2018enhancing, lee2018training, hsu2020generalized}. For example, recent work~\cite{hsu2020generalized} uses temperature scaling and an input preprocessing strategy to make the max class probability a more effective score for detecting out-of-distribution data. Such studies are complementary to SelfChecker: they could first check for the input being out-of-distribution, and then SelfChecker can check the prediction.
In addition, our problem cannot be subsumed by confidence calibration. As stated in ConfidNet~\cite{corbiere2019addressing}, confidence calibration helps to create confidence criteria but ConfidNet’s focus is failure prediction. Comparing SelfChecker against a technique with temperature scaling is inappropriate because using temperature scaling to mitigate confidence values doesn’t affect the ranking of the confidence score on different classes and therefore cannot separate errors from correct predictions.

In the SE community, several studies consider checking a DL model's trustworthiness in \emph{deployment}. \textsc{SelfOracle}, proposed by Stocco et al.~\cite{stocco2020misbehaviour}, estimates the confidence of self-driving car models. 
In their work, an alarm is triggered if the confidence of the model output is lower than a pre-defined threshold, in which case a human is then involved. It is designed for the scenario in which inputs are temporally ordered, such as video frames. Its performance is limited on other DNN types (see Section~\ref{sec:exp}). 
Wang et al.~\cite{wang2020dissector} propose \textsc{Dissector} to detect inputs that deviate from normal inputs. It trains several sub-models on top of the pre-trained DL model for validating samples fed into this DL model. But the generation of sub-models is manual and time-consuming, and \textsc{Dissector} does not provide an explicit design of the threshold for distinguishing inputs, which depends on the model and dataset.
In the DL community, researchers have developed new learning-based models to measure confidence \cite{lakshminarayanan2017simple,devries2018learning,papernot2018deep,jiang2018trust,corbiere2019addressing}. These models may also be untrustworthy and may suffer from, e.g., overfitting. In~\cite{papernot2018deep,jiang2018trust}, nearest-neighbor classifiers are built to measure the model confidence. A clear drawback of both approaches is the lack of scalability, since computing nearest neighbors in large datasets and complex models is expensive.
Corbi{\`e}re et al.~\cite{corbiere2019addressing} propose a new confidence model, namely ConfidNet, on top of the pre-trained model to learn the confidence criterion based on True Class Probability for failure prediction, which outperforms~\cite{jiang2018trust} in both effectiveness and efficiency. 
But its performance is limited due to overfitting since it is trained on the training dataset where there are few wrong predictions.
Except for~\cite{papernot2018deep}, which cannot scale to large datasets and models, none of the above papers provide alternative advice. In contrast, SelfChecker achieves both high alarm and advice accuracy (with sufficient validation data per class) using internal features extracted from the DNN.

\section{Limitations and conclusion}

\noindent \textbf{Limitations.} SelfChecker builds on an assumption
that the density functions and selected layers determined by the
training module can be used to check model consistency in
deployment. This assumption depends on whether the
training and validation datasets are representative of test
instances. SelfChecker is a layer-based approach that requires white-box access and will have more
limited power on shallow DNNs with few layers.

\noindent \textbf{Conclusion.} To be used in mission-critical
contexts, DNN outputs must be closely monitored since they will
inevitably make mistakes on certain inputs.

In this paper we hypothesized that features in internal layers of a
DNN can be used to construct a \emph{self-checking} system to check
DNN outputs. We presented the design of such a general-purpose system,
called SelfChecker, and evaluated it on four popular
publicly-available datasets (MNIST, FMNIST, CIFAR-10, CIFAR-100) and
three DNNs (ConvNet, VGG-16, ResNet-20).  SelfChecker produces
accurate alarms (accuracy of 60.56\%), and SelfChecker-generated
advice improves model accuracy on the 10-class dataset by 0.138\% on
average, within an acceptable deployment time (about
34.98ms). As compared to alternative approaches, SelfChecker achieves the highest F1-score with
68.07\%, which is 8.77\% higher than the next best approach
(ConfidNet).  In the self-driving car scenarios, SelfChecker triggers
more correct alarms than \textsc{SelfOracle} for both DAVE-2 and
Chauffeur models with a comparable number of false alarms.
SelfChecker is open source: \emph{https://github.com/self-checker/SelfChecker}.

\section*{Acknowledgment}
\xy{This work was supported in part by the National Research Foundation, Singapore and National University of Singapore through its National Satellite of Excellence in Trustworthy Software Systems (NSOE-TSS) office under the Trustworthy Software Systems – Core Technologies Grant (TSSCTG) award no. NSOE-TSS2019-05.}